\documentclass[10pt,journal,compsoc]{IEEEtran}
\usepackage{amssymb, amsmath, graphicx, paralist,subfigure,cite}
\usepackage{algorithm2e}
\usepackage{color}
\usepackage[table]{xcolor}
\newtheorem{lem}{Lemma}
\newtheorem{theorem}{Theorem}
\newtheorem{defn}{Definition}

\newtheorem{cor}{Corollary}

\def\mb{\mathbf}
\def\mc{\mathcal}

\begin{document}

\title{\huge{Observational Equivalence in System Estimation:\\ Contractions in Complex Networks}}

\author{Mohammadreza~Doostmohammadian,~\IEEEmembership{Member,~IEEE,}
        Hamid~R.~Rabiee,~\IEEEmembership{Senior~Member,~IEEE,}
        Houman~Zarrabi,~\IEEEmembership{Member,~IEEE,}
        and~Usman~Khan,~\IEEEmembership{Senior~Member,~IEEE}
\IEEEcompsocitemizethanks{\IEEEcompsocthanksitem M.~Doostmohammadian was with ICT Innovation Center for Advanced Information and Communication Technology, School of Computer Engineering, Sharif University of Technology, Tehran, Iran. He is currently with Mechanical Engineering Department, Semnan University, Semnan, Iran. \protect\\
E-mail: m.doostmohammadian@ictic.sharif.edu
\IEEEcompsocthanksitem H. R. Rabiee is with School of Computer Engineering, Sharif University of Technology, Tehran, Iran.\protect\\
E-mail: rabiee@sharif.edu
\IEEEcompsocthanksitem H. Zarrabi is with Iran Telecommunication Research Center (ITRC), Tehran,
Iran.\protect\\
E-mail: h.zarrabi@itrc.ac.ir
\IEEEcompsocthanksitem U. Khan is with Electrical and Computer Engineering Department, Tufts University, Medford, MA. \protect\\
E-mail: khan@ece.tufts.edu
}
\thanks{Manuscript received April 5, 2017.}}

\markboth{IEEE Transaction on Network Science and Engineering}%
{Doostmohammadian \MakeLowercase{\textit{et al.}}: Observational Equivalence in System Estimation: Contractions in Complex Networks}

\IEEEtitleabstractindextext{%
\begin{abstract}
	Observability of complex systems/networks is the focus of this paper, which is shown to be closely related to the concept of \textit{contraction}. Indeed, for observable network tracking it is necessary/sufficient to have one node in each contraction measured. Therefore, nodes in a contraction are equivalent to recover for loss of observability, implying that contraction size is a key factor for observability recovery. Here, using a polynomial order contraction detection algorithm, we analyze the distribution of contractions, studying its relation with key network properties. Our results show that contraction size is related to network clustering coefficient and degree heterogeneity. Particularly, in networks with power-law degree distribution, if the clustering coefficient is high there are less contractions with smaller size on average. The implication is that estimation/tracking of such systems requires less number of measurements, while their observational recovery is more restrictive in case of sensor failure. Further, in Small-World networks higher degree heterogeneity implies that there are more contractions with smaller size on average. Therefore, the estimation of representing system requires more measurements, and also the recovery of measurement failure is more limited. These results imply that one can tune the properties of synthetic networks to alleviate their estimation/observability recovery.
\end{abstract}

\begin{IEEEkeywords}
Observability, System Estimation, Contraction,  Complex network, Clustering Coefficient, Degree Heterogeneity
\end{IEEEkeywords}}

\maketitle
\IEEEdisplaynontitleabstractindextext
\IEEEpeerreviewmaketitle

\IEEEraisesectionheading{\section{Introduction}\label{sec_introduction}}

\IEEEPARstart{C}{omplex} networks have recently gained considerable attentions in control and estimation theory \cite{liu_pnas,ruths2014control,liu2016tutorial,das2015distributed,asilomar14,Liu_nature,barabasi2016social}. This interest stems from the challenge to understand and infer the fundamental aspects of system behavior. Such complex networks exist in nature for example in chemical reaction networks and biological networks \cite{liu_pnas} as in proteomics and gene networks.
Other than these natural complex networks, synthetic  large-scale networks are recently considered due to emergence of the so-called Internet-of-Things  (IoT) and Cyber-Physical-Systems (CPS) \cite{das2015distributed,asilomar14}. Interestingly the design of such man-made networks are significantly tied by control and estimation principles as they are genuinely constructed based on these principles \cite{liu2016tutorial}. Examples range from  consensus networks \cite{scientia} and social networks \cite{barabasi2016social} to more technological networks including electric power grids, computer networks, the Internet, etc.  Indeed, many networks are a formalism to describe phenomena and systems in real
life\footnote{In this paper a network describes the underlying dynamic system or phenomena. Therefore, throughout the paper the network and system are used interchangeably.}. In these and other similar applications the research focus is to uncover the tie between the internal system/network dynamics and the controllability and estimation properties.

It is known that, the internal states of complex systems are to a great extent dependent on each other, which is due to interaction of different  components on each other and therefore these complex systems are represented as networks \cite{liu_pnas}. This inter-dependence is such that by measuring and tracking certain variables of complex system one can infer sufficient information about the rest of the system for filtering and tracking purposes. This implies that  measuring well-selected variables give an \textit{observable} inference of complex system. The term observability is a measure defining whether the internal states of a system can be determined by knowledge of its measurements. The system is said to be observable if one can reconstruct the \textit{complete}  state of the complex system from the set of measured states also known as system outputs \cite{bay}.

There are different methods to check for observability of dynamic systems, namely: (i) algebraic method based on Gramian test \cite{bay}; (ii) the symbolic method also known as Popov-Belevitch-Hautus (PBH) test \cite{hautus}; and, (iii) the structural observability method introduced by pioneering work of Lin \cite{lin}. The first two methods are based on numerical values of system parameters while the third method is irrespective of the parameters and only relies on the \textit{structure} of the underlying system. Therefore, the third method has certain benefits over the two other methods as it is computationally efficient and only requires the structural information and sparsity pattern of the system instead of exact numerical values \cite{woude:03}. In other words, the structural method only relies on the knowledge of complex system as a graph/network and therefore is extensively studied in the literature \cite{lin,liu_pnas,ruths2014control,liu2016tutorial,asilomar14,Liu_nature,woude:03}.

The system graph representation, referred to as \textit{system graph}, is an abstract way of modeling complex systems and has recently applied widely in the literature to reduce the complexity of such systems \cite{liu_pnas,asilomar14,Liu_nature}.
In the system graph, every graph node represents a state (a variable or a parameter) and every link (or edge) between a pair of nodes represents derivative functional connection relating the state variables \cite{liu_pnas,Liu_nature,woude:03}. 
In general, graph representation approach is more conventional in networked systems \cite{acc13_mesbahi,jstsp14} where the network structure is embedded into the system structure. The structural observability of networks, or any system graph as a network, is also referred to as \textit{network observability} \cite{liu_pnas,Liu_nature} and is the adopted methodology in this paper.\footnote{See \cite{liu_pnas,Liu_nature,nonlin} for extension to nonlinear case.} 

The network observability, as an abstract observability model of the system is closely related to system graph properties.\footnote{It should be noted that structural observability and graph theoretic method  applied as a \textit{tool} to solve network observability problem. See reference \cite{liu_pnas,Liu_nature} for more information.} The main theorem on this topic is originally stated in \cite{lin} and further developed recently in \cite{liu_pnas,Liu_nature, asilomar11, commault_recovery, globalsip14, boukhobza_recovery,jstsp14}\footnote{Note that many of stated references deal with dual problem of network controllability. The graph properties and notions can be simply redefined for network observability.}. Existence of disjoint cycles and output connected paths in the graph is closely tied with its observability. In this direction, recently the concept of matching and dilations in graph \cite{Liu_nature}, and Strongly Connected Components (SCC) \cite{asilomar11} are introduced to be related to network observability/controllability. Among these, the concept of contraction is the focus of this paper. An introductory description of contraction in the network is the set of nodes contracting (linking) to fewer number of nodes. 

In system estimation perspective, nodes in the same contraction are observationally equivalent, i.e. in case of losing observability of one (unmatched) node/state in the network/system another node in the same contraction can be measured to recover for the loss of observability. This is applicable in estimation of large scale systems such as power grid \cite{asilomar14} and internet based autonomous systems. For example when a sensor fails to measure a state --due to excessive noise, disturbance, or even external attacks-- some necessary information of the system is lost and system/network cannot be tracked globally. To recover for this loss of information another sensor can be applied to measure equivalent state/node of the system/network. This is why the contractions play an important role in estimation. Indeed, one can apply a new sensor to measure an equivalent state in the same contraction and recover for the observability loss. In this regard, the size of contraction determines the possible number of equivalent options for observability recovery. Larger contractions imply more options among which one can choose the most efficient state measurement in terms of cost \cite{IJSS2017}, reliability, etc. This is the main motivation to analyze the size and distribution of contractions as they play a major role in system observability recovery.

\textit{Related Literature:} Structural observability of full-rank systems (having no contraction in system graph) is considered in \cite{liu_pnas,asilomar11}. In these works, structural observability is shown to be closely related to network SCC classification. In \cite{Liu_nature} using cavity method the authors find considerable relation between average network degree and number of unmatched nodes. As one of their main results, they find that denser networks have less number of unmatched nodes and therefore it is less challenging to control and direct the network to the desired state. In \cite{globalsip14} the authors consider \textit{distributed} estimation and  formulate necessary and sufficient conditions for distributed structural observability. This work finds the connection between the structure of complex system and the structure of monitoring sensor network.
In \cite{commault_recovery,boukhobza_recovery}, the authors classify sensors  based on their essentiality for observability using combinatorial algorithms with application to sensor  failure and diagnosis.
Among these and other literature, what missing is on the concept of contraction and the relation between distribution of contractions and properties of the network (or system graph).

\textit{Contribution:} In this paper, we study the properties of contractions in undirected networks/system-graphs as a key factor in estimation and observability. Adopting the structural/network observability method, the related question addressed here is that: how to find the equivalent state nodes in the network/system-graph to infer observationally equivalent information of the associated system? and we show that by finding contractions in the system-graph (or network), one can find the system states (or network nodes) equivalent in terms of observability and estimation. In this regard, the size of a contraction determines the potential number of \textit{equivalent} sensing locations in networks as model of complex systems, which is discussed in Section~\ref{sec_cont}. Further, a \textit{polynomial order} algorithm is applied to find the contractions in the system graph. This algorithm is a modification of the algorithm for unmatched node detection given in \cite{murota}. Contractions are of particular interest in recovering sensor failure and loss of observability in tracking/filtering  noise-corrupted global state of the system/network. Detailed discussion on application of contractions in system estimation and observability including example of observability recovery in Kalman filtering is provided in Section~\ref{sec_obsrv}.  Introducing the contraction set, the follow-up question is: how do the properties of these contraction sets change based on different characteristics of the underlying network? We investigate the effect of two factors on the size and distribution of contraction components: degree heterogeneity and clustering coefficient. First result of this paper is that the clustering coefficient as a network characteristic is related to average size and number of contractions. In particular, our results show that for Scale-Free networks, with power-law degree distribution, increase in clustering coefficient results in a decrease in average contraction size in the network. Further, we observe decrease in the number of contractions in high clustering coefficient Scale-Free networks. As the next contribution, we check the effect of degree heterogeneity in Small-World networks on contraction properties. Specifically, our results show that increase in randomness of link connectivity (tuning the $p$ factor) results in decrease in the average contraction size but increase in the number of contractions in the network. These results are addressed in Section~\ref{sec_rand}. Further in Section~\ref{sec_real}, as a practical contribution,  the contraction properties including the size distribution and prevalence are discussed for two real world networks: a Power-grid network and a Route-view network. Noting that the degree distribution of many real-world networks show power-law degree distribution, including the two example here, the results for these two practical examples corresponds with contraction properties of scale-free networks. More detailed discussion on these results and concluding remarks are stated in Section \ref{sec_conc}. It should be noted that in this paper the results are particularly stated for \textit{undirected} networks/system graphs.

\section{Notions on Graph Theory and Definition of Contraction} \label{sec_cont}
In this section we define the contraction sets in graphs by first introducing the relevant graph theoretic notions. Define a graph as~$\mc{G}=(\mc{V},\mc{E})$, where $\mc{V}$ is the node set containing $n$ graph nodes, and $\mc{E}=\{(v_i,v_j)\}$ is the set of edges connecting the nodes. Define a path as a sequence of distinct nodes with every consecutive nodes as  an edge in $\mc{E}$. Further, define a cycle as a path starting and ending at the same node. Define $\mc{N}(i)$ as the degree of node $i$.
The adjacency matrix of the graph $A_G=\{a_{ij}\}$ is defined as $a_{ij}=1$, if $(v_j,v_i) \in \mc{E}$, otherwise $a_{ij}=0$. We further introduce the following graph-theoretic concepts to define contractions:
\begin{itemize}
	\item \textit{Bipartite graphs:} Define a bipartite graph,~$\Gamma=(\mc{V}^+,\mc{V}^-,\mc{E}_\Gamma)$, such that its nodes are partitioned into two disjoint sets:~$\mc{V}^+$ and~$\mc{V}^-$, and all of its edges~$\mc{E}_\Gamma$ start in~$\mc{V}^+$ and end in~$\mc{V}^-$. We construct a bipartite graph,~$\Gamma$, from ~$\mc{G}$ with the edge set~$\mc{E}_{\Gamma}$, defined as the collection of~$(v_j^-,v_i^+)$, if ~$(v_j,v_i) \in \mc{E}$.\footnote{Note that, in general, edges in a bipartite graph have no direction. However in this paper, following the definition in \cite{murota}, it is assumed that the edges have direction from~$\mc{V}^+$ to $\mc{V}^-$. This kind of representation is later used in the definition of \textit{Auxiliary graph}.}
	
	\item \textit{Matching:} A matching,~$\underline{\mc{M}}$, on the system graph,~$\mc{G}$, is defined as a subset of the edge set,~$\mc{E}$, with no common end-nodes. In the bipartite graph,~$\Gamma$, it is defined as a subset of edges where no two of them are incident on the same vertex in $\mc{V}^+$, i.e. all the edges in~$\mc{M}$ are all disjoint. The number of edges,~$|\underline{\mc{M}}|$, is the size of the matching. A matching,~$\underline{\mc{M}}$, with maximum size is called maximum matching, denoted by~$\mc{M}$, which is non-unique in general.
	
	\item \textit{Matched/Unmatched nodes:} Let~$\mc{M}$ be a maximum matching defined on the bipartite graph,~$\Gamma$. Let~$\partial \mc{M}^+$ and~$\partial \mc{M}^-$ denote the nodes incident to~$\mc{M}$ in~$\mc{V}^+$ and~$\mc{V}^-$ respectively. Denote by~$\delta \mc{M}$ the set of unmatched nodes in~$\mc{V}^+$ as~$\delta \mc{M} = \mc{V}^+ \backslash \partial \mc{M}^+$. Note that maximum matching $\mc{M}$ is not unique in general.
	
	\item \textit{Auxiliary graph}, denoted by~$\Gamma^\mc{M}$, is a graph associated to a maximum matching,~$\mc{M}$. It is constructed by reversing all the edges of maximum matching,~$\mc{M}$, and keeping the direction of all other edges~$\mc{E}_{\Gamma} \backslash \mc{M}$, in the bipartite graph,~$\Gamma$. This graph is defined to localize the contractions in the system graph.
	
	\item \textit{${\mc{M}}$-alternating path:} In the auxiliary graph, define an ${\mc{M}}$-alternating path as a sequence of edges starting from an unmatched node in~$\delta \mc{M}$ and every second edge in~$\mc{M}$, and denote it by $\mc{Q}_{\mc{M}}$. The name comes from the alternating edges between unmatched part,~$\mc{E} \backslash \mc{M}$, and matched part,~$\mc{M}$, of the auxiliary graph. 	
	
	\item \textit{${\mc{M}}$-augmenting path:} In the auxiliary graph, define an ${\mc{M}}$-augmenting path, denoted by $\mc{P}_{\mc{M}}$, as an ${\mc{M}}$-alternating path with begin node and end node in $\delta \mc{M}$. 	
\end{itemize}
Having defined these preliminary notions on graph theory, the notion of a contraction set is defined as follows:
\begin{defn}
	In the auxiliary graph representation of a network,~$\Gamma^\mc{M} _A$, define a contraction set for every unmatched node~$v_j \in \delta \mc{M}$, as the set of nodes containing all states in $\mc{V}^+$ reachable by ${\mc{M}}$-alternating paths starting from~$v_j$. Denote this set by $\mc{C}_i$ and further define~$\mc{C}$ as the set of all contractions, i.e.~$\mc{C}=\{\mc{C}_1,...,\mc{C}_m\}$. Intuitively, in graph $\mc{G}$, a contraction set defines nodes that are connected (contracted) to less number of nodes.\footnote{It should be mentioned that the concept of contraction is dual of dilation defined in the network controllability problem \cite{Liu_nature}. In a dilation set, less number of nodes are dilated into more number of nodes. So that we don't need to continually refer to the dual graph, we define a contraction that is the natural dual of dilation.}
\end{defn}
\begin{figure}[t]
	\centering
	\includegraphics[width=3.5in]{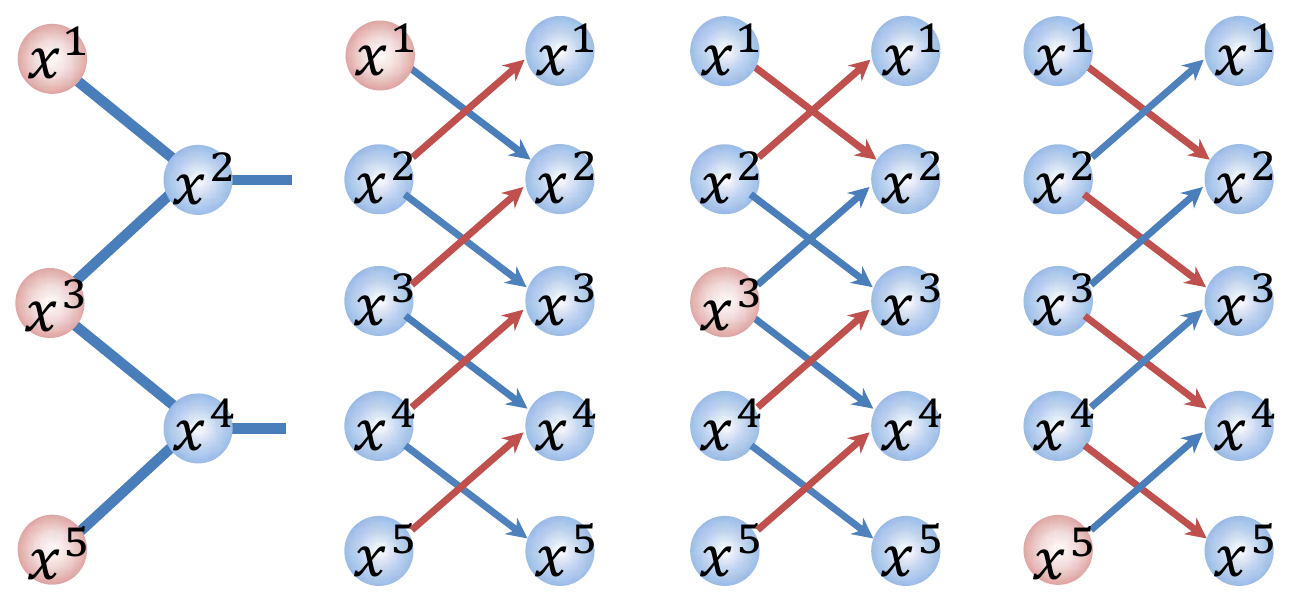}
	\caption{This figure shows a network contraction in the left, where the three contraction nodes are shown in red color. Figures in the right show three different  maximum matching in bipartite representations of the same contraction. The red edges represent maximum matching and the red node represents the unmatched node.}
	\label{fig_3nodecont}
\end{figure}
\begin{figure}[t]
	\centering
	\includegraphics[width=2.7in]{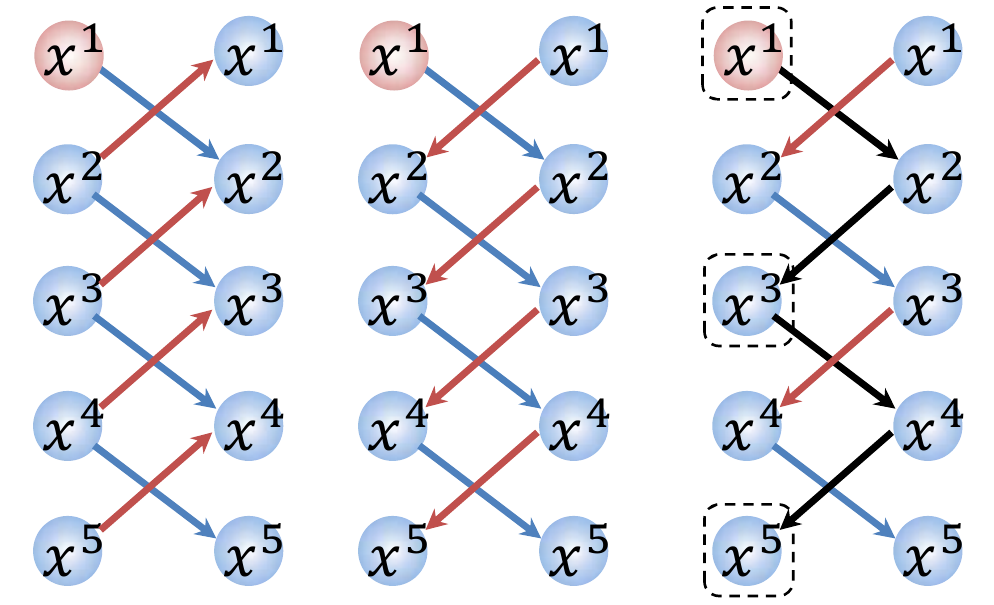}
	\caption{This figure illustrates the procedure of finding contractions explained in the paper. Graph in the left shows one possible matching and the unmatched node in the bipartite representation of the graph in Fig~\ref{fig_3nodecont}. The middle graph shows the auxiliary representation, where all matching edges are reversed. In the right graph an ${\mc{M}}$-alternating path is shown in black. Starting from the unmatched node, this path is used to find the contraction nodes (shown by dashed squares). Later in this paper we name these contraction nodes as observationally equivalent nodes.}
	\label{fig_3nodecont2}
\end{figure}
 For better illustration of the above definitions a contraction of 3 nodes into 2 nodes is shown in Fig.~\ref{fig_3nodecont}. The bipartite representation, the maximum matching, and the unmatched node are illustrated in the figure. We further illustrate the definition of auxiliary graph and ${\mc{M}}$-alternating path in Fig.~\ref{fig_3nodecont2}. The algorithm to find the contraction sets in network is given in Algorithm~\ref{alg_cont}. 

\begin{algorithm} \label{alg_cont}
	\textbf{Given:} System graph $\mc{G}_A$
	
	\KwResult{Contractions $\{\mc{C}_1,...,\mc{C}_m\}$}
	Construct the bipartite graph $\Gamma=(\mc{V}^+,\mc{V}^-,\mc{E}_\Gamma)$\;
	Find a matching $\underline{\mc{M}}$ as the set of edges with no common end nodes \;
	Construct the auxiliary graph $\Gamma^{\underline{\mc{M}}}_A$ by reversing the edges in matching $\underline{\mc{M}}$\;
	Define~$\partial {\underline{\mc{M}}}^+$ as the nodes in~$\mc{V}^+$  incident to~${\underline{\mc{M}}}$ \;
	Define  the set of unmatched nodes~$\delta{\underline{\mc{M}}}$ as $\delta {\underline{\mc{M}}} = \mc{V}^+ \backslash \partial {\underline{\mc{M}}}^+$\;
	Define the ${\mc{M}}$-alternating path,~$\mc{Q}_{\underline{\mc{M}}}$, as a sequence of edges starting from an unmatched node in~$\delta {\underline{\mc{M}}}$ and every second edge in~${\underline{\mc{M}}}$\;
	Define an ${\mc{M}}$-augmenting path,~$\mc{P}_{\underline{\mc{M}}}$, as an ${\mc{M}}$-alternating path with begin node and end node in $\delta {\underline{\mc{M}}}$\;
	\While{augmenting path $\mc{P}_{\underline{\mc{M}}}$ exist}{
		\For{nodes in $\delta \underline{\mc{M}}$}{
			Find $\mc{P}_{\underline{\mc{M}}}$ \;
			$\underline{\mc{M}} = \underline{\mc{M}} \oplus \mc{P}_{\underline{\mc{M}}}$ \;
		}
	}
	
	Construct the auxiliary graph $\Gamma^{\mc{M}}_A$ for the maximum matching ${\mc{M}}$\;
	Define ~$\partial {\mc{M}}^+$ for the maximum matching~${\mc{M}}$ and define $\delta {\mc{M}} = \mc{V}^+ \backslash \partial \mc{M}^+$\;
	Define the ${\mc{M}}$-alternating path,~$\mc{Q}_{\mc{M}}$, for the maximum matching~${\mc{M}}$\;	
	\For{nodes in $\delta \mc{M}$}{
		Find alternating paths $\mc{Q}_{\mc{M}}$ in $\Gamma^{\mc{M}}_A$ \;
		Put all nodes in $\mc{V}^+$ reachable by $\mc{Q}_{\mc{M}}$ in $\mc{C}_i$\;}
	
	\textbf{Return} $\mc{C}_i, i = \{1,...,m\}$\;\
	
	\caption{Contraction Detection Algorithm: the first loop in the algorithm gives the maximum matching,~$\mc{M}$, and unmatched nodes; the second loop in the algorithm gives the contraction set for each unmatched node.} 
\end{algorithm}
In Algorithm~\ref{alg_cont}, $\oplus$ is the XOR operator in set theory. As a result of this operator, each augmenting path increases the size of the matching  till it reaches the maximum matching. The computational complexity  of this algorithm is on the order of $\mc{O}(\sqrt{n} |\mc{E}|)$ or $\mc{O}(n^{\frac{5}{2}})$ in worst case.
In general, given the system graph~$\mc{G}_A$ there are other efficient algorithms to compute the maximum matching,~$\mc{M}$, e.g., the maximum flow algorithm~\cite{hopcraft}. The notions $\Gamma,\mc{M}$ can be obtained by the Dulmage-Mendelsohn (DM) decomposition~\cite{dulmage58}. 
Other than DM decomposition, maximum matchings can be efficiently computed in~$\mc{O}(\sqrt{n} |\mc{E}_A|)$ using the approach in~\cite{maxmatching}.
In the following, we state two main lemmas relating the maximum matchings and contractions.

\begin{lem} \label{lem_unmatched}
	Any choice of maximum matching,~$\mc{M}$, includes one and only one unmatched node in every contraction $\mc{C}_i, i\in \{1,...,m\}$.
\end{lem}
	The detailed proof is provided in \cite{murota,berge}.

\begin{lem} \label{lem_contr}
	For two sets of maximum matching,~$\mc{M}_1 \neq \mc{M}_2$, any unmatched  node~$v_i \in \delta \mc{M}_1$ can be reached along an alternating path from a  node~$v_j \in \delta \mc{M}_2$. This further implies that the set~$\mc{C}$ is the same for any choice of maximum matching.
\end{lem}
	The proof is given in \cite{murota}.

\section{Application in Observability and System Estimation}\label{sec_obsrv}
In this section, we first discuss the concept of structural observability in networks and then its application to system estimation. To further illustrate the results a network estimation example is provided.
\subsection{Network Observability} \label{subsec_netobsrv}
Observability of networks quantifies whether given measurements contain sufficient information to  comprehensively reconstruct the states of all nodes in the network. For a network, or a system graph representing a complex system, the necessary and sufficient conditions for (structural) observability is given in the following theorem.
\begin{theorem} \label{thm_obsrv}
	A network (or system graph) is structurally observable if and only if:
	(i) every node can reach to an output/measurement via a path of state nodes, and (ii) there exist a family of disjoint cycles and output-connected paths covering all nodes.
\end{theorem}
	The proof is given in \cite{lin} and in \cite{rein_book} for the dual case of structural controllability.
In Theorem~\ref{thm_obsrv}, condition (i) is known as \textit{accessibility} and condition (ii)
as the \textit{S-rank} condition. Note that for connected undirected networks the accessibility is already satisfied. This is because, in a connected undirected  network every node is reachable by every other node and therefore output connectivity of one node implies the reachability of all other nodes to that output.

\begin{theorem} \label{thm_unmatched}
	In a connected undirected  network with the set of unmatched nodes, $\delta \mc{M}$, observation/measurement of every unmatched node is necessary and sufficient for network observability.
\end{theorem}
	The proof is given in \cite{Liu_nature} for the dual case of network controllability.

Following the definition of contraction and results in previous section here we state the theorem on the concept of observational equivalence in contractions.
\begin{theorem} \label{thm_contr}
	In a connected undirected  network with the set of contractions $\mc{C}=\{\mc{C}_1,...,\mc{C}_m\}$, a measurement/observation of one state node in every contraction $\mc{C}_i, i\in \{1,...,m\}$ is necessary and sufficient for network observability.
\end{theorem}
	From Theorem~\ref{thm_unmatched} observation of every unmatched node is necessary and sufficient for network observability. Note that based on Lemma~\ref{lem_unmatched} for every contraction $\mc{C}_i$, every node $v_j$ is an unmatched node for a choice of maximum matching $\mc{M}$. This implies that observing at least one node in every contraction is necessary for observability. Further, by measuring node $v_j$ in $\mc{C}_i$ from Lemma~\ref{lem_contr} all other nodes in $\mc{C}_i \backslash v_j$ are matched for the choice of maximum matching $\mc{M}$ and therefore only one node is sufficient for network observability.

\begin{lem} \label{lem_rankdeficiency}
	Number of contractions in a network $\mc{G}$ equals the (structural) rank deficiency of its associated adjacency matrix, $A_G$. Indeed the rank deficiency of the adjacency matrix, $A_G$, equals  the number of unmatched nodes in the network $\mc{G}$.
\end{lem}
	Indeed from Lemma~\ref{lem_unmatched} the number of contractions equals the number of unmatched nodes in the network. Note that, by definition, unmatched nodes appear on acyclic part of network while the cyclic part is completely matched. It is known that the rank of the network adjacency matrix is structurally defined by the number of nodes included in a set of disjoint cycles \cite{godsil}. This implies that the rank deficiency can be structurally defined by the number of unmatched nodes, which are contained in the acyclic part of the network.

The concept of contraction is closely related to the concept of \textit{observational equivalence}. Let $C_i$ denote the measurement matrix of node/state $x^i$. Let $\mc{O}(A,C_i)$ represent the observability Grammian of the pair $A$ and $C_i$. The observational equivalence relation among two states/nodes $x^i$ and $x^j$, denoted by $x^i \sim x^j$, is defined as:
\begin{eqnarray}
	\mbox{rank}~\mc{O}(A, C_i) = \mbox{rank}~\mc{O}(A, C_j) = \mbox{rank}~\mc{O}\left(A, \left[
	\begin{array}{c}
		C_i\\
		C_j
	\end{array}
	\right]
	\right)
\end{eqnarray}
Note that this follows the three properties of the equivalence relation, i.e. transitivity, reflexivity, and symmetry.\footnote{Transitivity implies that if $x^i \sim x^j$ and $x^j \sim x^k$, then $x^i \sim x^k$. Reflexivity implies that every state/node is equivalent to itself, and symmetry implies that  $x^i \sim x^j$, then  $x^j \sim x^i$.}
\begin{lem} \label{lem_equiv}
	The algebraic implication of observational equivalence relation among states in each contraction is defined as follows. For any two (or more) measurements of states $x^i$ and $x^j$ in the same contraction, the structural-rank recovery of system matrix $A$ is equal to $1$, i.e.,
	\begin{eqnarray} \label{eq_srank}
	\mbox{S-rank}\left(
	\left[
	\begin{array}{c}
	A \\
	C_{i}
	\end{array}
	\right]
	\right) &= &
	\mbox{S-rank}\left(
	\left[
	\begin{array}{c}
	A \\
	C_{j}
	\end{array}
	\right]
	\right)
	\nonumber \\
	&=& \mbox{S-rank}\left(
	\left[
	\begin{array}{c}
	A \\
	C_{i} \\
	C_{j}
	\end{array}
	\right]
	\right)
	\nonumber \\
	&=& \mbox{S-rank}(A)+1.
	\end{eqnarray}
	where S-rank  implies the structural-rank\footnote{Note that, the structural rank (or S-rank) is defined as the maximum rank of the system matrix,~$A$, by changing its free parametric entries. In the system graph,~$\mc{G}$,~S-rank is the size of the \textit{maximum matching},~$\mc{M}$, see~\cite{shields,van1999generic} for details.} of the matrix.
\end{lem}
The proof directly follows the three properties of observational equivalence relation. One can easily check that the reflexivity, symmetry, and transitivity of equation \eqref{eq_srank} directly follows. 
\begin{cor} \label{cor_equiv}
	Theorem~\ref{thm_contr} along with Lemma~\ref{lem_unmatched},~\ref{lem_contr},~and \ref{lem_equiv} imply that all nodes in a contraction are \textit{equivalent} 
	in terms of observability. In other words, measurement of \textit{any} node in each contraction, assuming that all other contractions have one observation, provides network observability.
	As a result of the equivalent observability relation, nodes in the same contraction recover loss of observability.
\end{cor}

In other words, in the case of observation failure of a node, say node $v_i$, some information of the system is lost. In this case, observation of another node, say node $v_j$, sharing a contraction with node $v_i$ recovers the  observability loss. In this regard, the \textit{size of contraction} defines the possible number of equivalent sensing nodes for recovering observability loss.The implication of equivalent relation is further discussed in next subsection; we show how dynamic systems can be represented structurally as networks, where we can apply the above Theorems and Lemmas to find equivalent states in terms of system observability and estimation.

\subsection{System Estimation}
Consider the system model to be a discrete-time linear dynamic system\footnote{The results carry forward are also applicable to continuous-time systems.}:
\begin{eqnarray}\label{sys1}
\mb{x}_{k+1} &=& A\mb{x}_k + \mb{v}_k,
\end{eqnarray}
with ~$\mb{x}_k\in\mathbb{R}^n$
\begin{eqnarray} \nonumber
\mb{x}_k =
\left(
\begin{array}{c}
x_k^1\\
\vdots\\
x_k^n
\end{array}
\right)
\end{eqnarray}
as the state vector, ~$A=\{a_{ij}\}\in\mathbb{R}^{n\times n}$ as the system matrix, and~$\mb{v}_k\sim\mathcal{N}(0,V)$ as the system noise. Assume the dynamical system to be monitored by measurement/observation model:
\begin{eqnarray}\label{sys2}
\mb{y}_k = C\mb{x}_k + \mb{r}_k,
\end{eqnarray}
where
\begin{eqnarray}\nonumber
\mb{y}_k =
\left(
\begin{array}{c}
y_k^1\\
\vdots\\
y_k^m
\end{array}
\right),~~
C =
\left(
\begin{array}{c}
C_1\\
\vdots\\
C_m
\end{array}
\right),~~
\mb{r}_k =
\left(
\begin{array}{c}
r_k^1\\
\vdots\\
r_k^m
\end{array}
\right),
\end{eqnarray}
Here, $\mb{r}_k\sim\mathcal{N}(0,R)$ is the observation noise
with~$R = \mbox{blockdiag}[R_1,\ldots,R_N]$, and~$C$ is
the measurement matrix.

In structured systems theory, the LTI system in Eqs.~\eqref{sys1}-\eqref{sys2}, can be modeled as a \emph{system graph}. In this scenario, every node is a system state and every edge represents the interaction of two states based on the system matrix,~$A$. Denote the set of system states by~$\mc{X}\triangleq\{x^1,\ldots,x^n\}$ and denote the set of system observations/measurements by~$\mc{Y}\triangleq\{y^1,\ldots,y^m\}$. Then the system graph is defined by~$\mc{G}_A=(\mc{X},\mc{E}_A)$ where the edge set,~$\mc{E}_A$, is defined as~$\mc{E}_A=\{(x^i,x^j)~|~a_{ji}\neq0\}$, to be interpreted as~$x^i\rightarrow x^j$. One should note that, in this graph representation of  system the structure of system graph only relies on free parametric entries of matrix $A$. In other words, the graph structure depends on each entry $a_{ij}$ being a free parameter and not on the exact numerical value of $a_{ij}$. Therefore, any Linear Structure Invariant (LSI) system with fixed structure and time-varying parameters can be modeled as a system graph.\footnote{This is not a straighforward procedure as the edge weights vary over time while the structure is time-invariant. Note that here we only convey the idea behind LSI dynamics with fixed sparsity pattern on the adjacency matrix.} 
The motivation of applying graph representation of system is that one can check its characteristics by using equivalent graphical properties. The system characteristic of interest here is system observability, which plays a crucial role in system estimation and filtering. To illustrate more we consider the role of system graph observability in Kalman estimation as discussed next.

Let~$\widehat{\mb{x}}_{k|k}$ be the Kalman estimator
tracking system state $\mb{x}_k$  at time~$k$ given all the measurements, $\mb{y}_k$, up to time~$k$. The dynamics of this estimator is defined as follows \cite{kalman:61}:
\begin{eqnarray}\label{eq_kalman1}
\widehat{\mb{x}}_{k|k-1} &=& A\widehat{\mb{x}}_{k-1|k-1}
\\ \label{eq_kalman2}
\widehat{\mb{x}}_{k|k} &=&\widehat{\mb{x}}_{k|k-1} + K_k C^T (\mb{y}_k-C\widehat{\mb{x}}_{k|k-1})
\end{eqnarray}
where the~$K_k$ is the Kalman gain computed in a recursive procedure as proposed by Kalman \cite{kalman:61}. It can be shown that the error, $\widehat{\mb{e}}_{k|k} = \mb{x}_k - \widehat{\mb{x}}_{k|k}$, in this estimator is given by,
\begin{eqnarray}\label{ge}
\widehat{\mb{e}}_{k|k} = (A - K_kC^TCA)\widehat{\mb{e}}_{k-1|k-1} + \eta_k,
\end{eqnarray}
where the vector~$\eta_k$ collects the remaining terms (noise terms) that are independent of $\widehat{\mb{e}}_{k-1|k-1}$ and $\widehat{\mb{e}}_{k|k}$. It is known that the
dynamics of Kalman error,~$\widehat{\mb{e}}_{k|k}$ is stable if the measurements defined by matrix $C$ give an observable inference of system defined by $A$ \cite{kalman:61}. In other words, the Mean Squared Estimation Error (MSEE) reaches bounded stability over time if the pair $(A,C)$ is observable. Following the results of the graph-theoretic method in Section~\ref{subsec_netobsrv}, we consider two applications in the following.
\begin{enumerate} [(i)]
	\item As the first application one can check the observability constraint using results of Theorem~\ref{thm_contr}. For a system to be observable, according to Theorem~\ref{thm_contr}, one state node in each contraction set in the system graph has to be measured.
	
	\item The other, and more important, application is in case of observability loss. Assume that one (or more) of the sensor measurements fail and therefore the system is not observable anymore. To recover for this loss of observability, one can assign measurements of equivalent states as stated in  Corollary~\ref{cor_equiv}. The set of equivalent states for observability and estimation can be determined by finding contractions in the system digraph representation using Algorithm~\ref{alg_cont}. For example, loosing the measurement of state $x^i$ one can measure another state $x^j$ sharing a contraction with $x^i$ in $\mc{G}_A$ to recover for system observability.
\end{enumerate}
These graph-theoretic applications are  explained more in the following example.

\begin{figure}[hbpt!]
	\centering
	\includegraphics[width=2.45in]{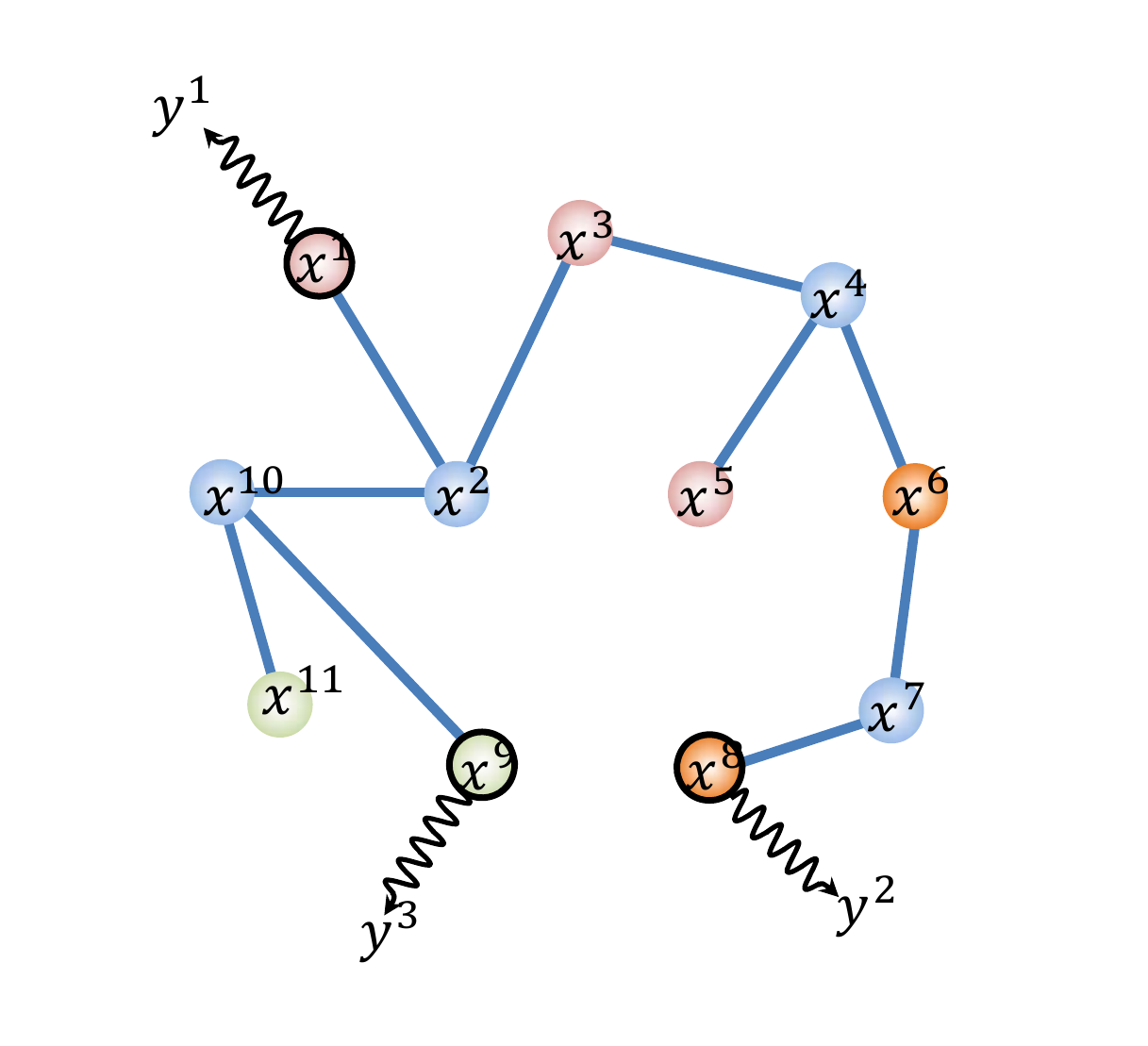}
	\includegraphics[width=2.45in]{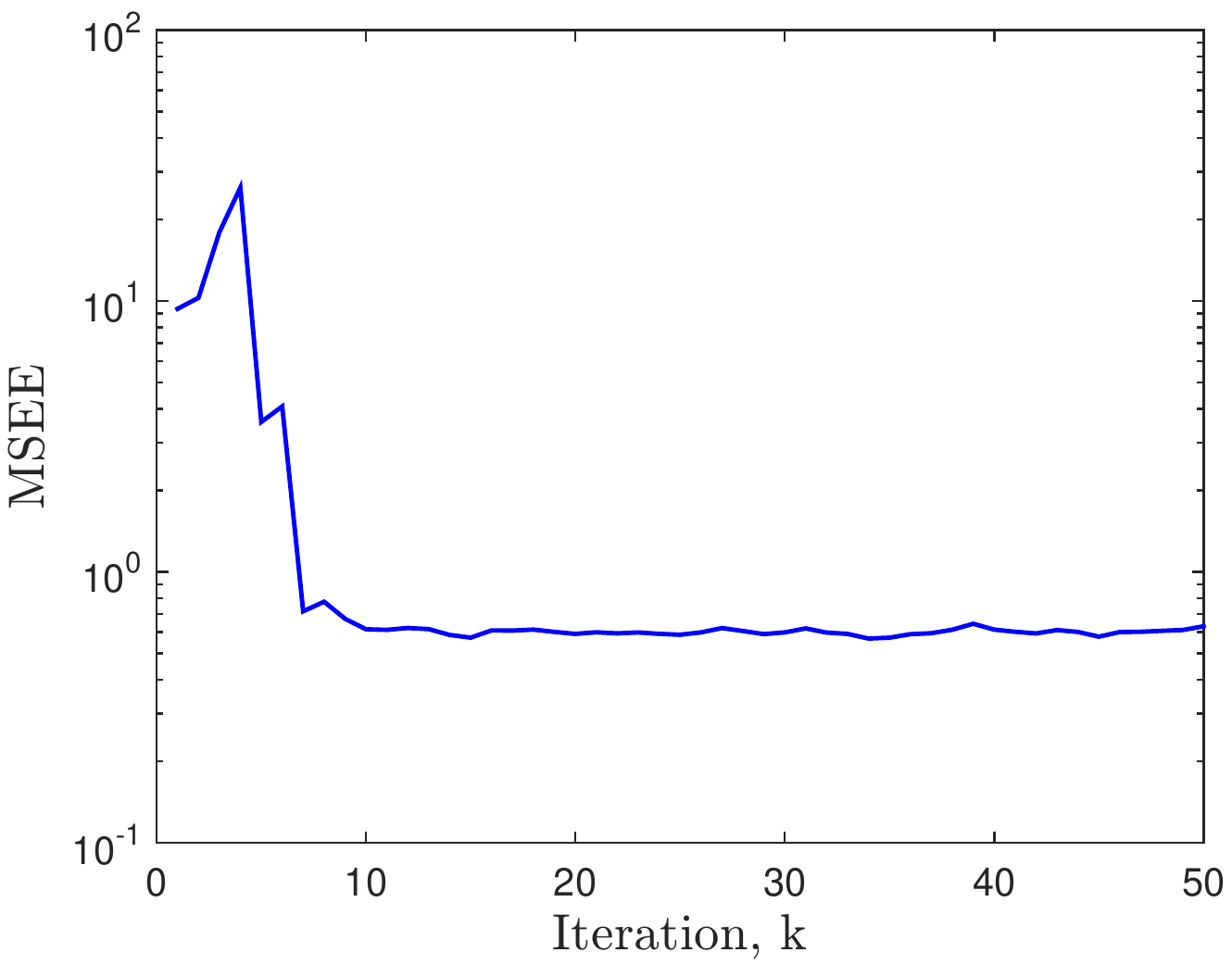}
	\caption{(Top) This figure shows the example system graph with $3$ measurements of states $x^1,x^8,x^9$. The same-colored state nodes in red, green, and orange each represent states in the same contraction component and the blue states are not part of any contraction. (Bottom)  The time evolution of the MSEE using the estimator in Eqs.~\eqref{eq_kalman1}-\eqref{eq_kalman2} applied on the same system. The three measurements give an observable inference and therefore the MSEE is bounded steady-state stable.}
	\label{fig_graph1}
\end{figure}

\textit{Illustrative example:} Consider a system with $n=11$ states represented as a system graph in Fig.~\ref{fig_graph1}-Top.
\begin{figure}[hbpt!]
	\centering
	\includegraphics[width=2.45in]{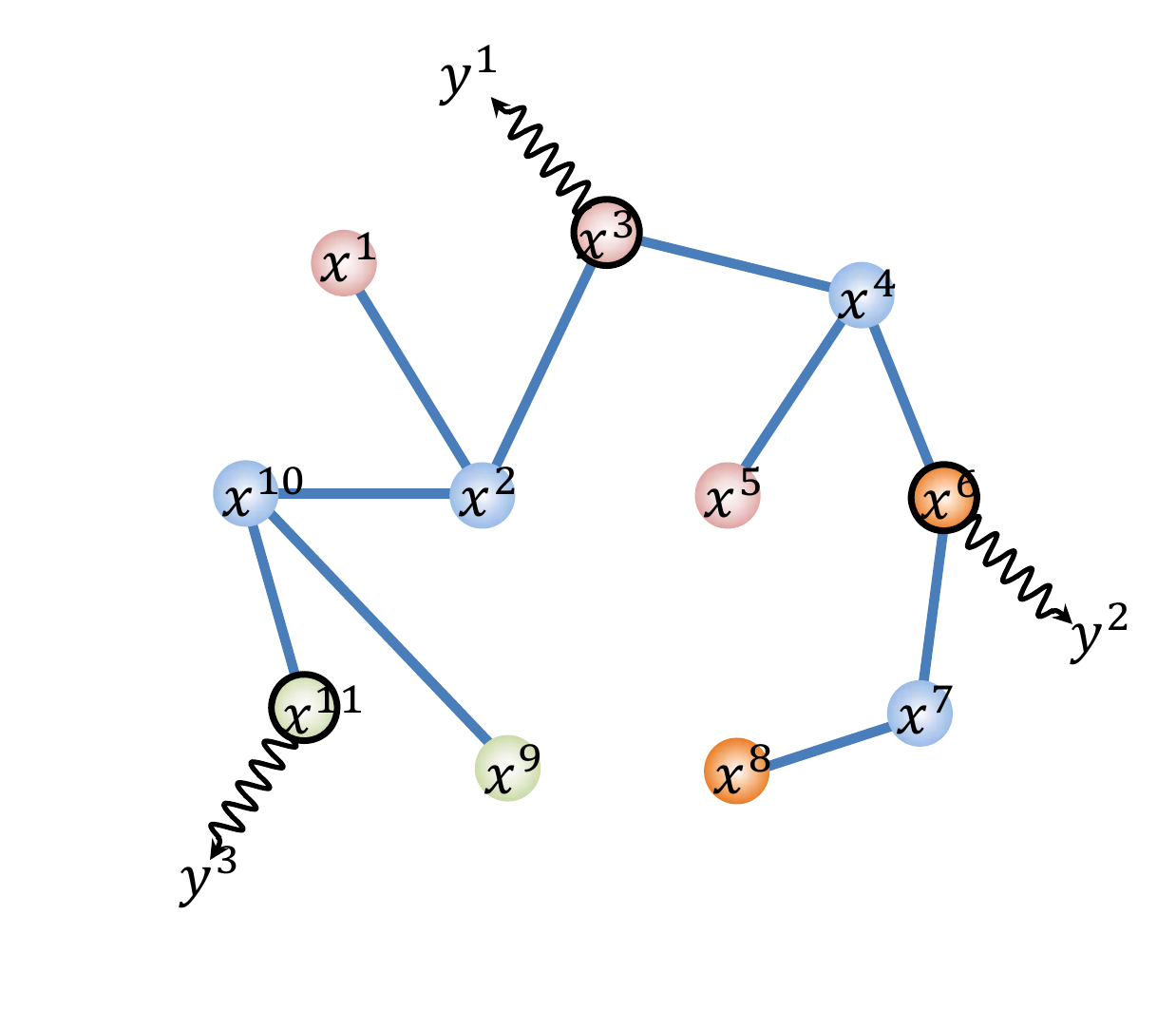}
	\includegraphics[width=2.45in]{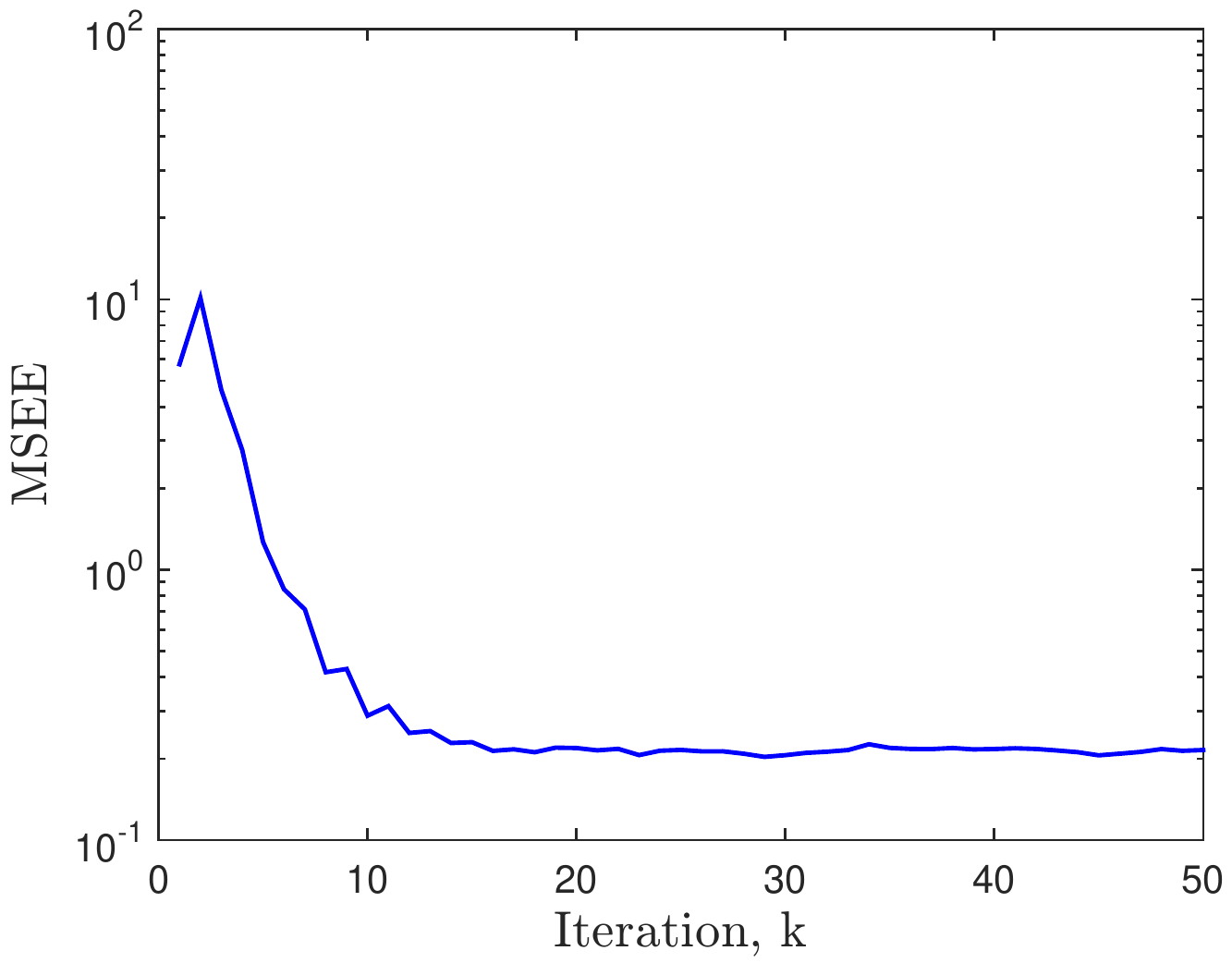}
	\caption{(Top) This figure shows the same example system graph in Fig.~\ref{fig_graph1} with $3$ new measurements of states $x^3,x^6,x^{11}$ sharing a contraction with the states measured in Fig.~\ref{fig_graph1}. (Bottom)  The time evolution of the MSEE of system estimation using the estimator in Eqs.~\eqref{eq_kalman1}-\eqref{eq_kalman2}. As it can be seen, the equivalent measurements also provide an observable estimation with bounded steady-state  MSEE.}
	\label{fig_graph2}
\end{figure}
Assume this graph represents a dynamic system, where each node is system state and each link represents the dynamic interaction of two states (for more details on such representation of systems as networks see \cite{Liu_nature}). The associated system matrix elements  in $A_G$ (i.e. the link weights) are chosen randomly. For sake of illustration and avoiding trivial solutions the elements in $A_G$ are such that the spectral radius of adjacency matrix is greater than $1$, $\rho (A_G)>1$, i.e. the system is unstable. 
To determine the necessary states for observability we find the set of contractions in the system graph using Algorithm~\ref{alg_cont} as $\mc{C}_1=\{x^1,x^3,x^5\}$, $\mc{C}_2=\{x^6,x^8\}$, and $\mc{C}_3=\{x^9,x^{11}\}$.\footnote{Note that contraction $\mc{C}_1$ is similar to the contraction described in Fig.~\ref{fig_3nodecont} and~\ref{fig_3nodecont2}.} The system is tracked by $m=3$ measurement of three states each in one contraction set of the system graph. This satisfies the condition in Theorem~\ref{thm_contr} for observability and  thus leads to stable estimation. These measurements along with system parameters are used in estimator Eqs.~\eqref{eq_kalman1}-\eqref{eq_kalman2} to estimate and track the global state $\mb{x}$  of the system over time iterations $k$. The Mean Squared Estimation Error (MSEE) over time is shown in Fig.~\ref{fig_graph1}-Bottom, which is bounded. Note that if we loose the measurement of a state in a contraction, according to Theorem~\ref{thm_contr} we loose the system observability. Without observability, we loose the stability of the MSEE and the estimation error goes unbounded. To recover for loss of observability, one can take a measurement of an equivalent state in the same contraction, as shown in Fig.~\ref{fig_graph2}-Top. Indeed, measuring any state in the same contraction set is sufficient for observability and yields stable estimation as shown in Fig.~\ref{fig_graph2}-Bottom. The key point is that number of possible state to recover for observability directly relates to the size of contraction sets. In this example, there are two options to recover for loss of observation of $x^1$, while there is only one replacement for observation of $x^8$ and  replacement for observation of $x^9$.

\section{Synthetic and Real Case Studies} \label{sec_sim}
In this section, we analyze the number and size of contraction sets in both real and random complex networks. Recall that  the contraction size is of interest because it determines the number of equivalent nodes for observability recovery, and number of contractions determine  the number of node measurements necessary and sufficient for observability. First, random networks as models of complex systems are reviewed and relation between  features of these networks and size/number of contractions are discussed. Next, the distribution of contraction sets in some examples of real-world networks are analyzed. Here, we proceed by first reviewing the  definitions of relevant network properties.
\begin{defn}
	The local clustering coefficient of a node in a graph is defined as the fraction of pair of node neighbors that are linked together. On the other hand, the global clustering coefficient is defined as the fraction of the closed triplets (triangles) to the total number of the triplet paths in the graph \cite{newman2003structure}.
	Mathematically the clustering coefficient is defined as:
	\begin{eqnarray}
	CC(i) = \frac{2.tr(i)}{\mc{N}(i)(\mc{N}(i)-1)}
	\end{eqnarray}
	where $tr(i)$ is the number of triangles that node $i$ forms with two of its neighbors. The global clustering coefficient of the network is defined as,
	\begin{eqnarray}
	CC = \frac{3.tr}{trp}
	\end{eqnarray}
	where the $tr$ is the number of triangles and $trp$ is the number of connected triplets in the network. It is known that the clustering coefficient is a good measure of well-connectivity of the network and presence of strong community-structure in the network \cite{alstott2016cluster,newman2003structure}.
\end{defn}


\begin{defn} Degree heterogeneity is an intuitive concept related to the degree distribution of networks. Degree heterogeneity, as opposed to degree homogeneity, determines if the nodes in the network have various degrees (heterogeneous), or have similar degrees (homogeneous) to one another.	
Various measures of degree heterogeneity resembling the global differences in the node degrees are discussed in \cite{wu2010heterogeneity,jacob2017measure}. 
	The most well-known formula for degree heterogeneity is given by the variance of node degrees as follows \cite{jacob2017measure}:
	\begin{eqnarray}
	VAR = \frac{1}{n}\sum_{i=1}^{n}(\mc{N}(i) - \bar{\mc{N}})^2
	\end{eqnarray}	
	where $\bar{\mc{N}}$ is the average node degree,
	\begin{eqnarray}
	\bar{\mc{N}} = \frac{1}{n}\sum_{i=1}^{n}\mc{N}(i)
	\end{eqnarray}	

\end{defn}

\noindent More details on these definitions can be found in \cite{newman2003structure,jacob2017measure}.

\subsection{Contraction sets in Random Networks} \label{sec_rand}
Random graphs are widely used to model complex systems facilitating analysis of different processes over networks, e.g., spreading processes or cascading failures \cite{newman2003structure,barabasi2003scale,barabasi_albert1999,Holme2002clusteringScaleFree,Toivonen2006social,watts1998smallworld,newman2002random}. The graphs are called random since the nodes in the graph are randomly connected with each other. We investigate two well-known models for random graphs. We particularly  analyze the relation between number and size of contractions with clustering coefficient in Scale Free networks and with degree heterogeneity in Small-World Networks as discussed next.

\textit{Scale-Free networks:} Many complex networks are modeled by this type of random network. It is known that degree distribution of such networks follows a power-law distribution \cite{barabasi2003scale}, i.e. 
the portion of nodes having degree $d$, represented by $f(d)$, follows the following formula:
\begin{eqnarray}
f(d) = d^{-\sigma},~2<\sigma <3
\end{eqnarray}
In log-log scale, the distribution represents a linear function, hence it is named Scale-Free (SF) network. This implies that these graphs have a large number of low-degree nodes and few hubs with high connectivity. A well-known approach to build such networks is proposed by Barabasi and Albert \cite{barabasi_albert1999}. The Barabasi and Albert (BA) approach considers an initial graph of few number of nodes, called \textit{initial seed} where recursively a new node with $m$ new links is added to the network. The probability that the new node makes a link to old nodes is proportional to the degree of old nodes, implying that the new node \textit{preferably} links to high degree nodes, and is known as \textit{preferential attachment}. In this method hubs with high degree are more likely to connect to the newly added nodes while the low degree nodes are unlikely to gain new links. These types of Scale-Free networks, e.g. BA model, are known to have low clustering coefficient. Therefore recently new random models of networks are proposed in the literature to account for high clustering of real networks \cite{Holme2002clusteringScaleFree,Toivonen2006social}. These works propose to modify preferential attachment method such that the resulting networks, beside having power-law distribution, have high clustering. 
The network growth procedure is similar to the preferential attachment of BA model with some modification. Similar to BA model, they consider an initial seed. Then, recursively a new node connects to $m_r$ nodes in the network based on preferential attachment. But further, in each step the new node also makes $m_s$ links to randomly chosen neighbors of preferentially attached nodes in the network. This additional step is called \textit{triad formation}. This increases the prevalence of triads (triangle cliques) in the network, and therefore results in high clustering coefficient. These random networks are called Clustered Scale Free (CSF) networks.

\begin{figure}[hbpt!]
	\centering
	\includegraphics[width=3in]{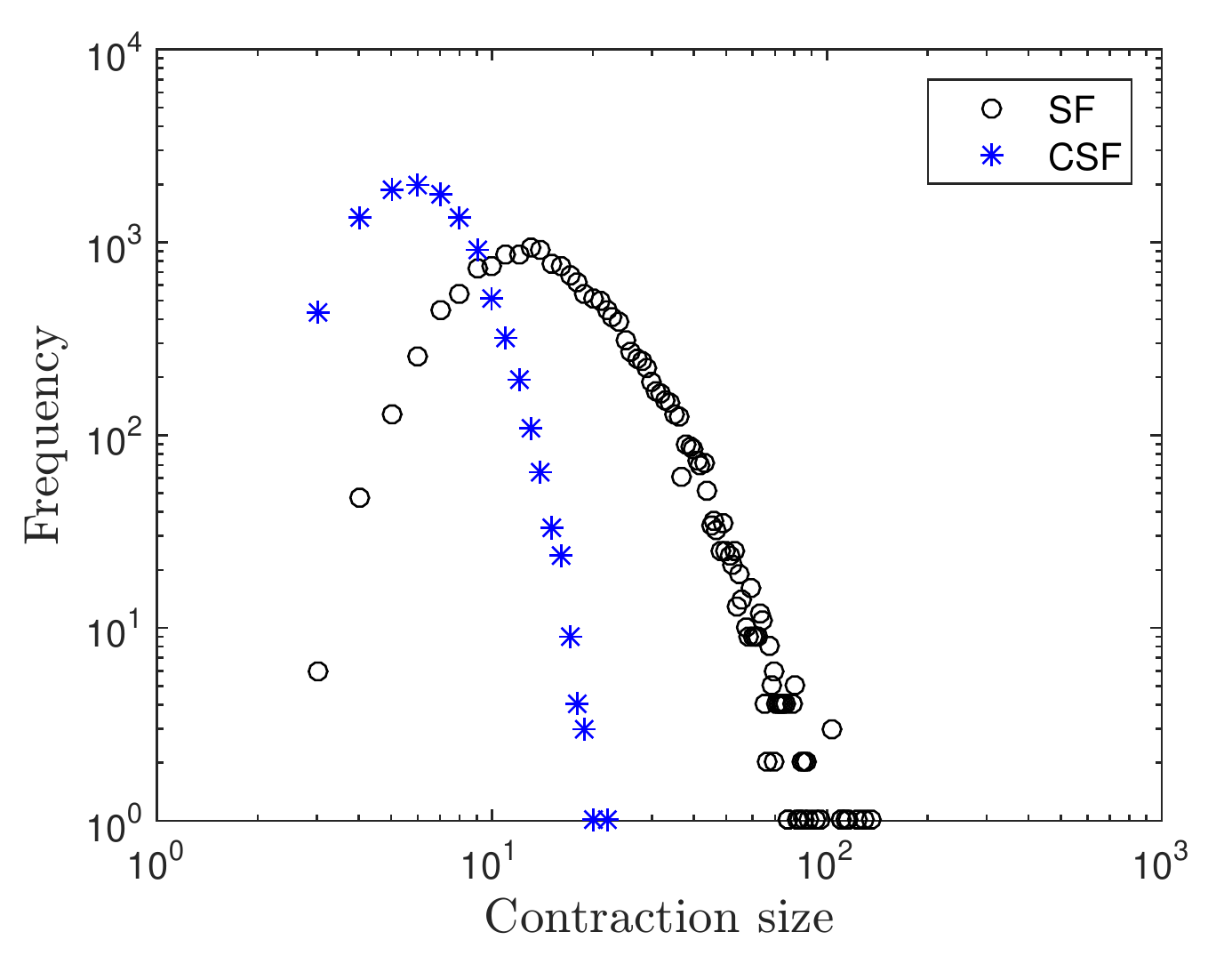}
	\caption{This figure shows the distribution of contraction size on SF and CSF networks with similar number of preferentially attached nodes. The simulation is performed over $1000$ realizations of $1000$-nodes networks.}
	\label{fig_SF_CSF}
\end{figure}

For these networks we analyze the size and number of contractions. It should be noted that for simulation we consider  $m=m_r+m_s$, i.e.  number of links each new node makes in SF network equals number of links each new node makes in CSF network. This implies that the number of edges in both SF and CSF network are the same. This also implies that the average degree of the network is equal for both SF and CSF networks. This is important as all properties of both SF and CSF networks are similar except their clustering coefficient \cite{Holme2002clusteringScaleFree,Toivonen2006social,assenza2008enhancement}. Simulations are performed over $1000$ different realizations of 1000 node SF and CSF networks and the results are shown in Fig.~\ref{fig_SF_CSF}.
As it can be seen contractions are more prevalent in SF networks as compared to CSF networks and there are more contractions (and unmatched nodes) in SF networks as compared to CSF networks. Further, the contraction sets are in average larger in SF as compared to CSF networks. The results for $1000$ network realizations are summarized  in Table~\ref{tab_SF}.
\begin{table}[hbpt!]
	\centering
	\caption{ Average size and number of contractions  in $1000$ realizations of SF and CSF networks.}
	\begin{tabular}{|l|c|c|}
		\hline
		Type of network&~ SF &~CSF\\
		\hline
		Average Contraction size & ~$18.89$  &~$6.72$ \\
		\hline
		Average number of Contractions & ~$156$  &~$109$ \\
		\hline
		\hline
	\end{tabular}
	\label{tab_SF}
\end{table}
This implies that by increasing the clustering coefficient as in CSF networks the number and size of contractions decreases.

\textit{Small-World networks:}
The idea behind this model is to imitate the graphical properties of real-world  networks. One of the main structural feature of the real graphs is that they show high level of community structure while keeping small average distance (shortest path), which is known as the small-world phenomena. Such features are not present in typical random models, for example in Erdos-Renyi graphs. Therefore, Watts and Strogatz \cite{watts1998smallworld} proposed a new semi-random graph, named Small-World model. The Watts and Strogatz (WS) model starts with a $k$-regular network in which every node is connected to its $k$ nearest neighbors (in both sides). Randomly pick links in the $k$-regular network with uniform probability $p$ independent of each other. Then, choose the end node of this link and randomly rewire it to another node. The rewiring must be such that the new link is not a self-link or a link that already exist in the network. By increasing the rewiring probability $p$ one can generate random networks which are more random in terms of their degree heterogeneity and as $p \rightarrow 1$ the model reaches an Erdos-Renyi (ER) graph with Poisson degree distribution \cite{newman2002random}. On the other hand, small $p$ implies that network conserves its regularity and degrees of most nodes lie around the average degree $2k$. Such networks have adjustable degree heterogeneity by tuning $p$.
Indeed, regular networks are the most degree homogeneous and next are small world networks with tunable degree heterogeneity by factor $p$. By increasing the factor $p$ the degree heterogeneity increases up to the point where $p=100\%$ and the graph models the ER network.

To relate the contraction size with degree heterogeneity, {$1000$ different realizations of 1000 node SW networks with $9$ different $p$ factors are considered and the simulation results are shown in Fig.~\ref{fig_SW}. Note that for this simulation only rewiring probability $p$ is changing, therefore  graph properties such as number of edges and average node-degree remains unchanged and the only property that changes is degree-heterogeneity \cite{dwivedi2017optimization}. The average size and number of contractions in $1000$ network realizations are also summarized  in Table~\ref{tab_SW}.
\begin{figure}[hbpt!]
	\centering
	\includegraphics[width=3in]{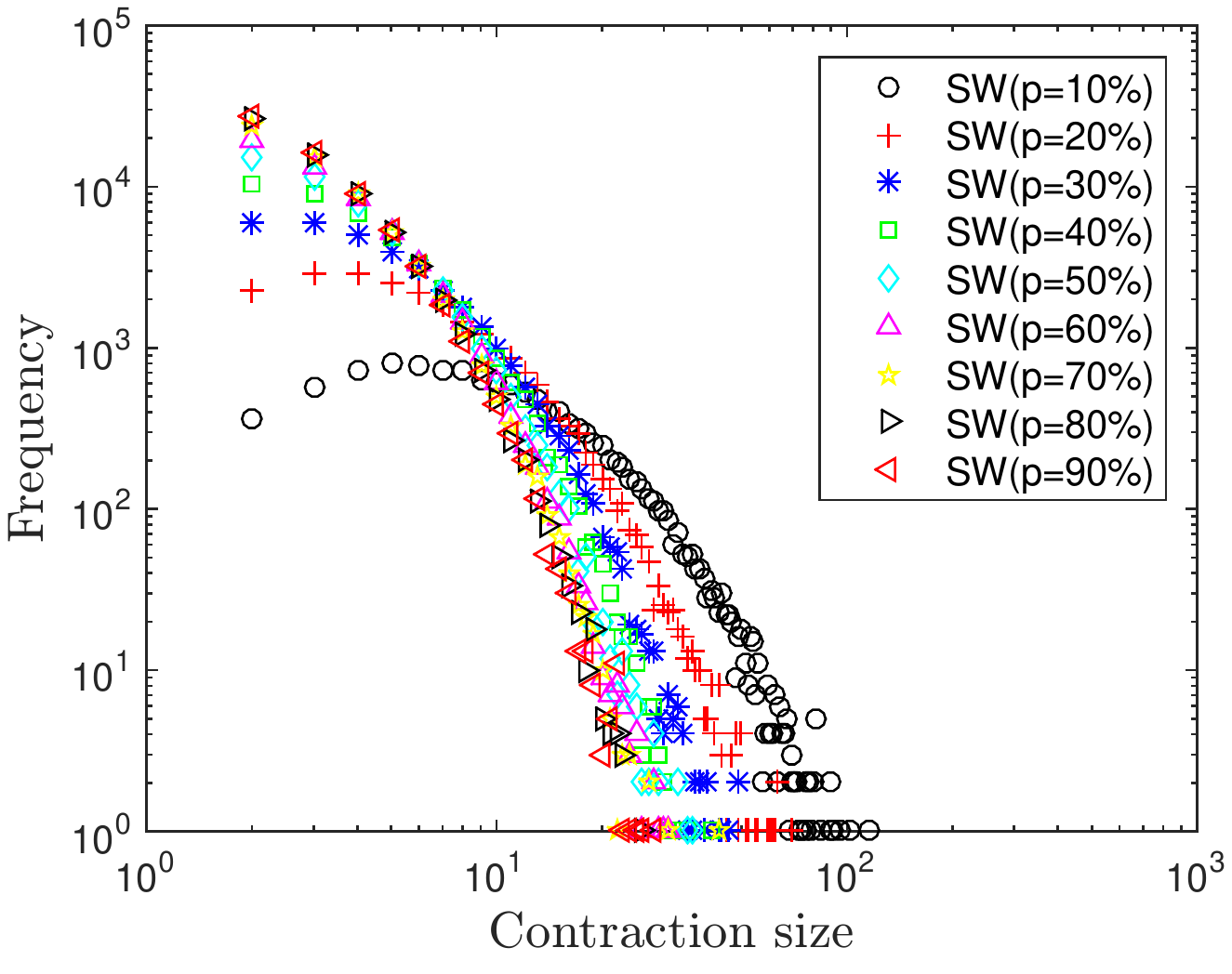}
	\caption{This figure shows the distribution of contraction size on different Small-World networks tuning the $p$ factor. The simulation is performed over  $1000$ realizations of $1000$-nodes networks.}
	\label{fig_SW}
\end{figure}
}
\begin{table}[hbpt!]
	\centering
	\caption{ Average size and number of contractions in  $1000$ realizations of SW networks with $10$ different  $p$ factors.}
	 \begin{tabular}{|l|c|c|c|c|c|}
		\hline
		$p$ factor of SW network& $10\%$ &$20\%$ &$30\%$ &$40\%$ &$50\%$\\
		\hline
	    avg Contraction size & $13.71$  &$7.62$ &$5.62$ &$4.68$ &$4.14$ \\
		\hline
		number of Contractions & $25$  &$47$ &$68$ &$86$ &$101$ 		\\
		\hline		
		\hline
		$p$ factor of SW network& $60\%$ &$70\%$ &$80\%$ &$90\%$ &$100\%$\\
		\hline
		avg Contraction size & $3.83$  &$3.61$ &$3.50$ &$3.44$ &$3.42$ \\
		\hline
		number of Contractions & $113$  &$123$ &$130$ &$132$ &$133$ 		\\
		\hline		
		\hline
	\end{tabular}
	\label{tab_SW}
\end{table}
As it can be seen from the results by increasing the $p$ factor and degree heterogeneity, in  average contractions are decreased in size but increased in number.

\subsection{Contraction sets in Real Networks} \label{sec_real}
\begin{figure}[hbpt!]
	\centering
	\includegraphics[width=3in]{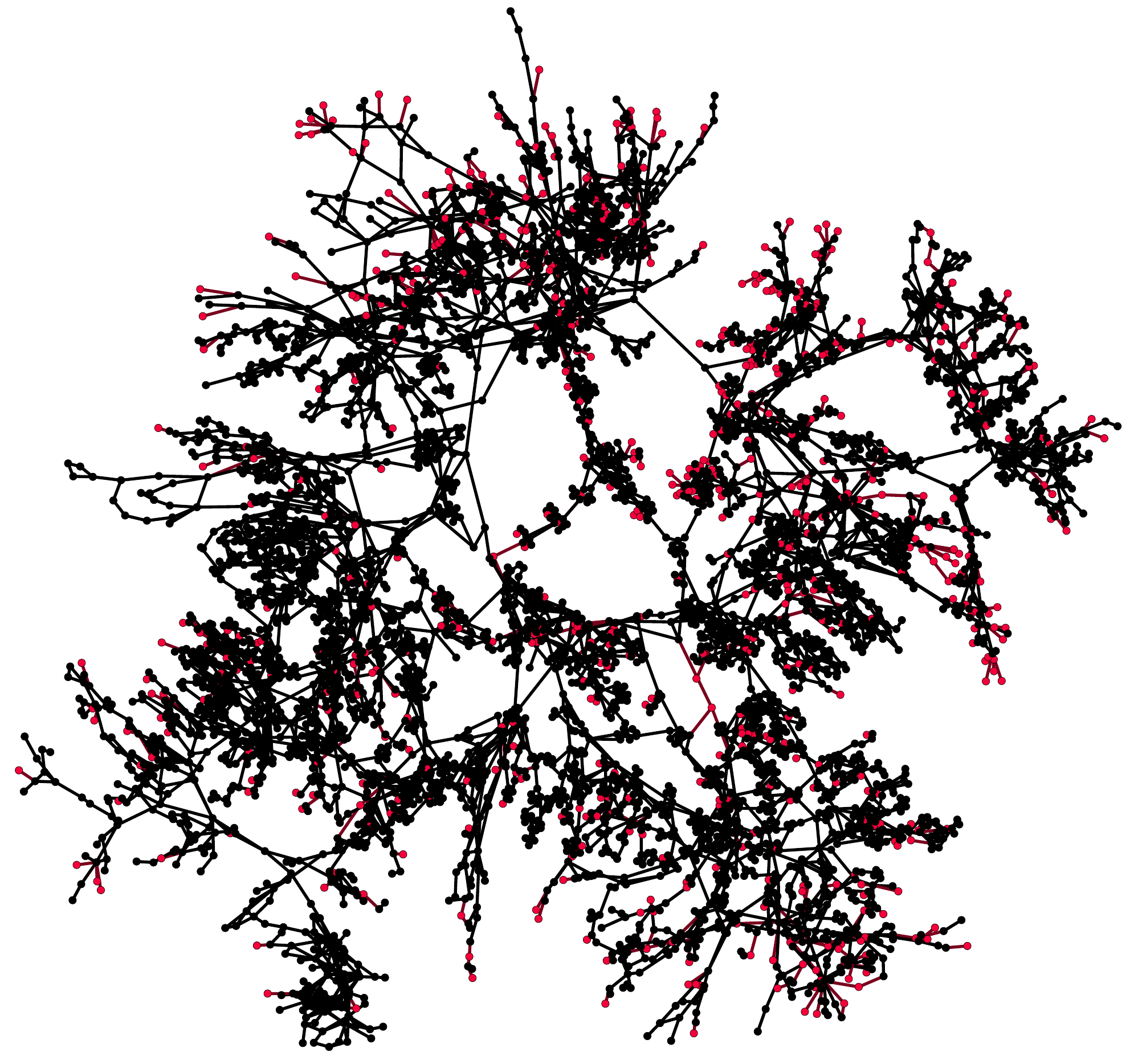}
	\caption{This figure shows the structure of Western-State Power grid network with $4941$ state nodes and $6594$ links; red nodes in the network represent unmatched nodes monitored by a sensor.}
	\label{fig_power}
\end{figure}

\textit{Power grid network:} As the first example, we consider the power grid network that represents the grid of the Western States of the United States of America.  In this network a link is a power supply line and a node represents either a generator, a transformator or a substation. The network is originally addressed in \cite{watts1998smallworld} but the data is taken from \cite{UCI_power} where the description of state nodes can be found. It is known that such networks resemble the sparsity of system dynamic matrix where the states represent power flow, voltage, or phase angles \cite{asilomar14,poor2012grid}, and therefore the network can model a dynamic system type of Eq. \eqref{sys1}, for more details see \cite{poor2012grid,khan2013secure,camsap11}. This network contains $6594$ interaction links connecting $4941$ state nodes.
Applying the DM decomposition finds one possible set of unmatched nodes in the network. The network structure is shown in Fig.~\ref{fig_power}, including $575$ unmatched nodes represented in red color. Recall that From Theorem~\ref{thm_unmatched}, for observable estimation all the unmatched nodes must be observed by a sensor. These set of observable measurements gives one possible stable estimation of system state nodes over time.

\begin{figure}[hbpt!]
	\centering
	\includegraphics[width=2.5in]{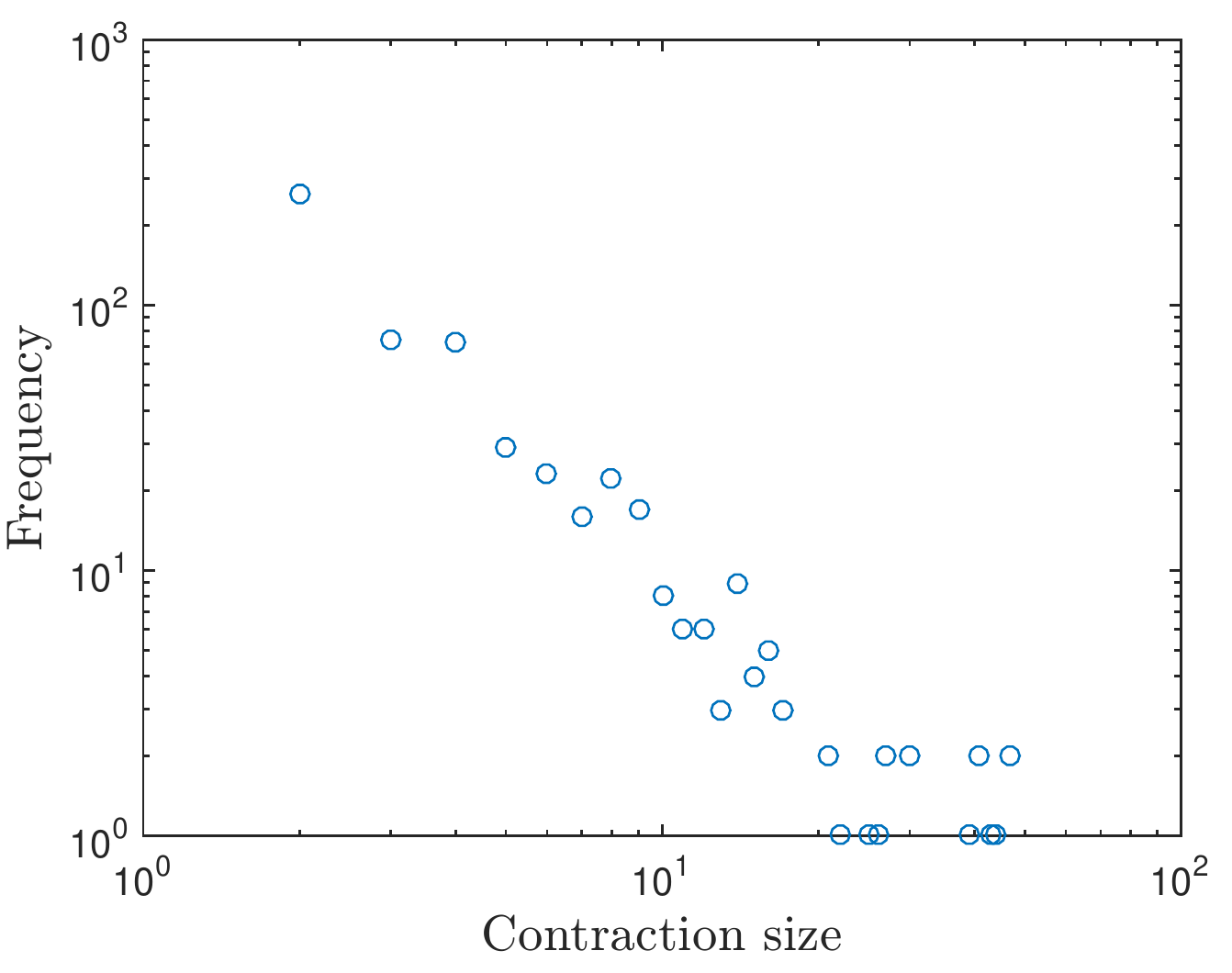}
	\caption{This figure shows the frequency of size of different contractions  in power grid network.} 	 \label{fig_power_distribution}
\end{figure}

\begin{figure}[hbpt!]
	\centering
	\includegraphics[width=2in]{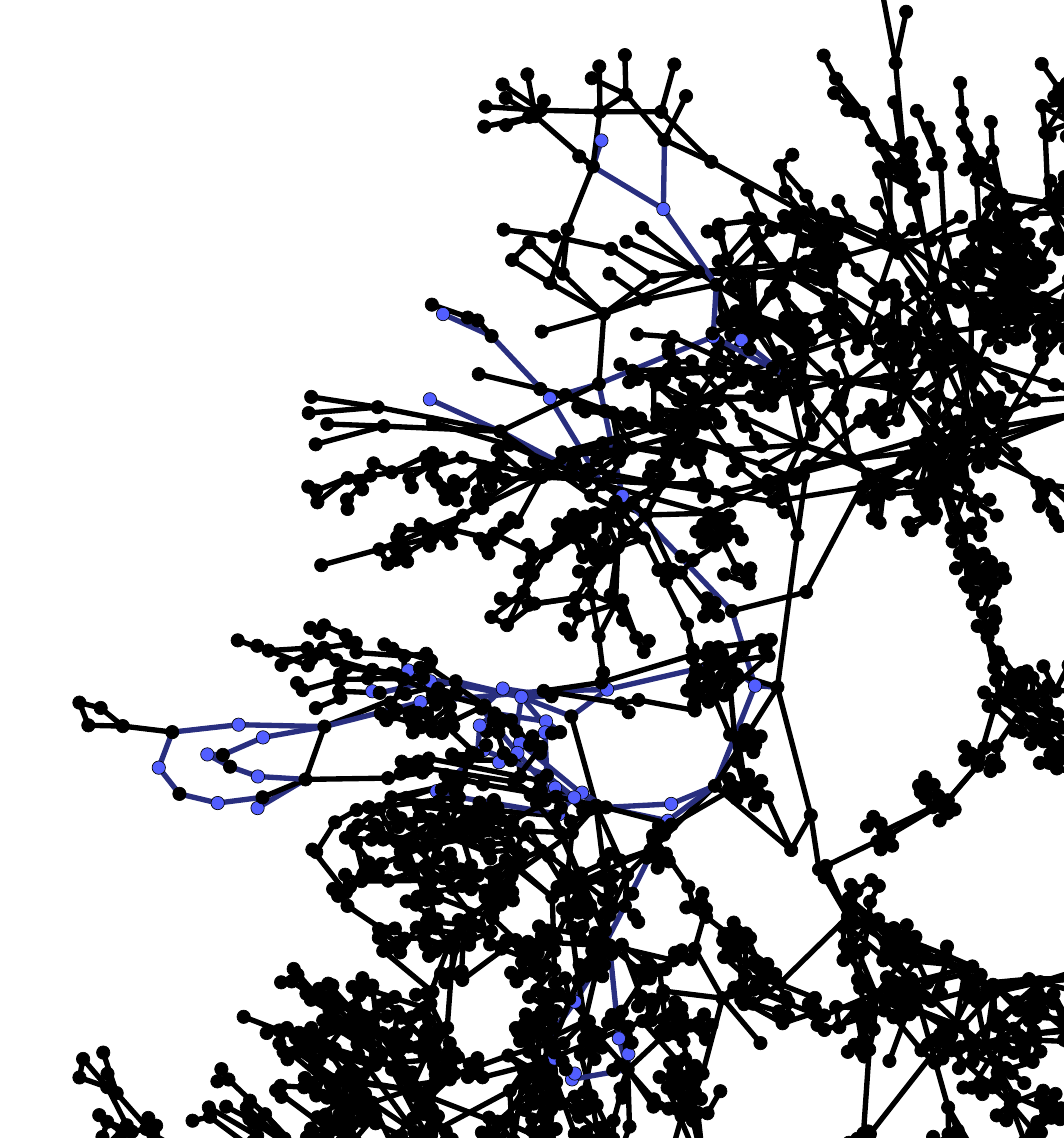}
	\includegraphics[width=1in]{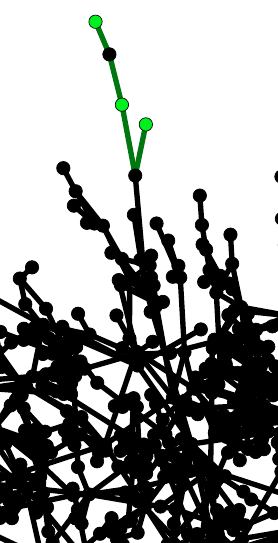}
	\caption{This figure includes two subnetworks of Power-grid network of Fig.~\ref{fig_power}. Each subnetwork shows an example of nodes making a contraction, represented as blue and green colored nodes. These colored state nodes are equivalent in terms of network observability.}
	\label{fig_power_cont}
\end{figure}

The distribution of all $575$ contractions in this network are shown in Fig.~\ref{fig_power_distribution}. It should be noted that the clustering coefficient of this network is $10.3\%$ and the average contraction size is $4.98$.
Applying the Contraction Detection Algorithm~\ref{alg_cont} finds all the contraction sets in the network, where two examples are shown in Fig.~\ref{fig_power_cont}. These examples include contraction sets of size $3$ (green colored contraction) and $52$ (blue colored contraction). Recall that from Corollary~\ref{cor_equiv} each contraction set associated with an unmatched node represents all possible options to recover for loss of measurement/observation. In this scenario, losing the observation of any node in the 3-nodes green contraction implies that there are measurements of only $2$ other nodes to recover for possible loss of observability, while for the blue colored contraction there are $51$ possible options to recover for the loss of observability.

\begin{figure}[hbpt!]
	\centering
	\includegraphics[width=3.3in]{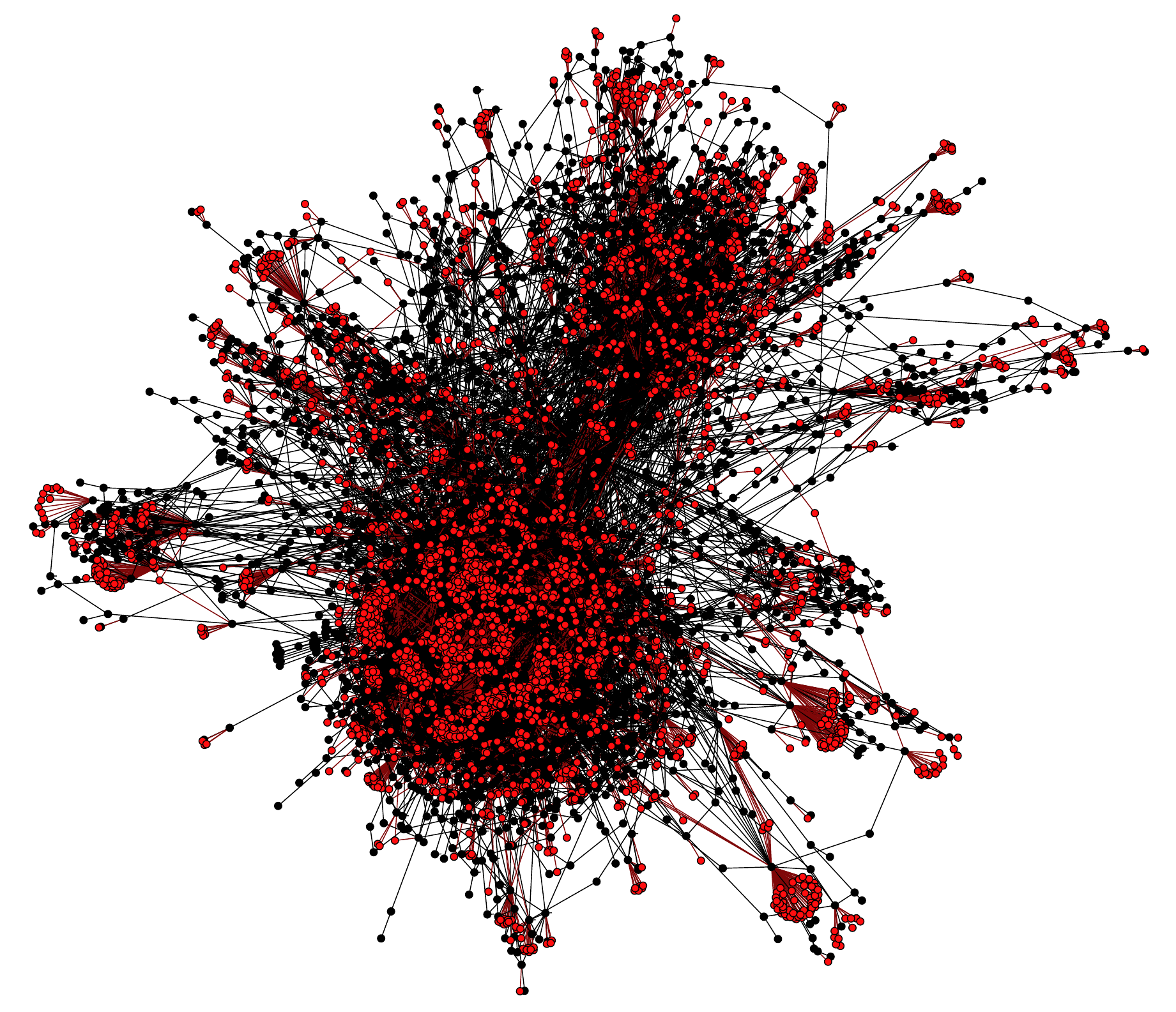}
	\caption{This figure shows the Route-views network representing internet-connected autonomous systems. The network contains $6474$ nodes connected with $13895$ links. Nodes in red color are unmatched nodes each monitored by a sensor.}
	\label{fig_routeview_unmatched}
\end{figure}

\textit{Route-view network:} This network represents the network of connected autonomous systems of Internet. Every node is an autonomous system and every link represents communication between two systems. The data is taken from \cite{Konect_Routeviews}, but the original description of the network is given in \cite{leskovec2007graph}. As stated in \cite{leskovec2007graph} every node is indeed a subgraph of Internet-connected routers that exchanges traffic flow with its peer neighbors. The network contains $6474$ nodes connected with $13895$ links, and is represented in Fig.~\ref{fig_routeview_unmatched}. In this figure regular nodes are represented in black while $3568$ unmatched nodes are shown in red color.

\begin{figure}[hbpt!]
	\centering
	\includegraphics[width=2.5in]{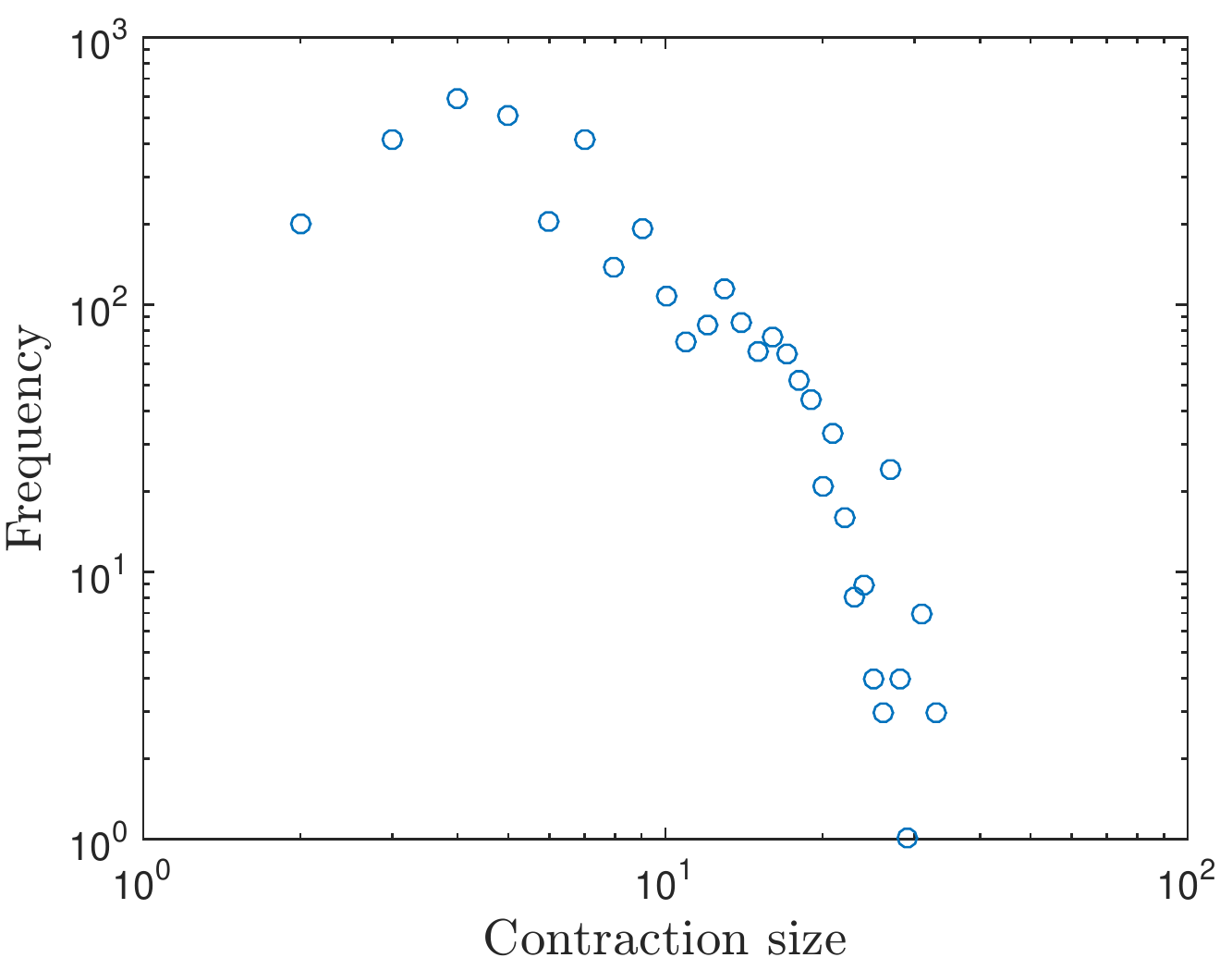}
	\caption{This figure shows the frequency of size of different contractions in Route-view network.}
	\label{fig_RouteView_distribution}
\end{figure}

Applying the Contraction Detection Algorithm~\ref{alg_cont} all $3568$ contractions in the network are found. The distribution of contraction sets is as shown in Fig.~\ref{fig_RouteView_distribution}.  The average size of contractions in this network is $7.65$ and the clustering coefficient of the network is $0.959 \%$.

\begin{figure}[hbpt!]
	\centering
	\includegraphics[width=2in]{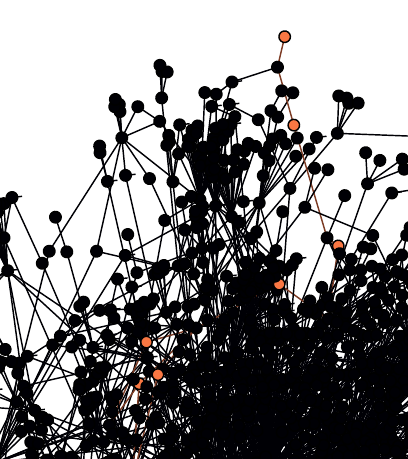}
	\includegraphics[width=1in]{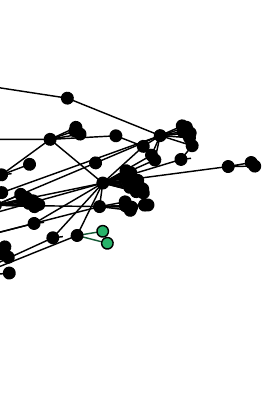}
	\caption{This figure shows two subnetworks of Route-view network of Fig.~\ref{fig_routeview_unmatched}. In each subnetwork colored nodes in orange and green represent example of nodes making a contraction. In network observability, these colored nodes represent equivalent states.}
	\label{fig_Routeview_cont}
\end{figure}

Two examples of contraction sets in the Route-view network are shown in Fig.~\ref{fig_Routeview_cont}; one includes set of $2$ contraction nodes (in green color) and the other one includes set of $7$ contraction nodes (in orange color).  As mentioned before, each contraction set represents the state nodes giving equivalent information for network observability and estimation.

The results for these two networks are summarized in  Table~\ref{tab_realnet}.
\begin{table}[hbpt!]
	\centering
	\caption{ Characteristics of two examples of real networks including average contraction size, ratio of number of contractions to number of nodes, and clustering coefficient.}
	\begin{tabular}{|l|c|c|}
		\hline
		Name of network&~ Power grid &~Route-view\\
		\hline
		avg Contraction size & ~$4.98$  &~$7.65$ \\
		\hline
		Contractions/nodes & ~$575/4941$  &~$3568/6474$ \\
		\hline
		Clustering Coefficient & ~$10.3\%$  &~$0.959\%$	\\
		\hline
		\hline	
	\end{tabular}
	\label{tab_realnet}
\end{table}

\section{Discussion and Conclusions} \label{sec_conc}

Comparing the SF and CSF network, we observe a significant raise in  average size of contractions in SF network. Noting that SF and CSF networks apply the same preferential  attachment model and are similar in terms of most graph statistics including power-law degree distribution and logarithmically increasing average shortest-path length \cite{Holme2002clusteringScaleFree,Toivonen2006social,assenza2008enhancement}, therefore, the only difference is low clustering coefficient as the key factor affecting the jump in average size of contractions in SF network. Similar statement holds for the average number of contractions in the network. Note that this number is decreased for clustered version of Scale-Free network while other network characteristics are unchanged. This implies that by increase in the clustering coefficient in average more contractions with larger size appear. This is also the case in real-world network examples stated in Section~\ref{sec_real}\footnote{Note that it is known that most real-world networks including the two examples given in this paper follow a power-law degree distribution \cite{barabasi_albert1999}.}. For Power grid network with high clustering coefficient the ratio of number of contractions to the total number of nodes is lower than the Route-view network with low clustering coefficient. Similar statement holds for the average contraction size as the size of contractions are in average smaller in Power grid network.
For observability and estimation of networks  with power-law degree distribution (SF and CSF networks) these results imply that: (i) estimation/tracking of such networks with high clustering coefficient  requires (in average) lower number of observations/measurements as there are less number of contractions, but (ii) in case of measurement/sensor failure there are less number of possible equivalent states for observability recovery as the average size of contractions are low. One application of these results is that one can tune the clustering coefficient of (synthetic) networks \cite{serrano2005tuning} to reduce  the challenge for observability recovery and estimation.

The other result of this paper is that in Small World networks the average size of contractions is to a great extent related to the degree homogeneity. Increasing the heterogeneity in Small World networks, by increasing the rewiring probability $p$ \cite{watts1998smallworld}, is one key factor on the decrease of average contraction size as mentioned in Table~\ref{tab_SW}. Note that by only changing the rewiring probability in SW networks the number of links, average node degree, and the size of graphs are unchanged. On the other hand, by increasing the degree heterogeneity, while the other graph characteristics of in SW networks are unchanged, the average number of contractions is increased. In terms of observability and estimation of SW networks these results imply that: (i) estimation of networks with high level of degree heterogeneity requires more number of measurements/observations which is due to prevalence of contractions, and (ii) in case of sensor/observation failure  there are less number of possible options to recover for the loss of observability as the contraction sets in average are smaller in degree heterogeneous SW networks. As an application of these results one may decrease the degree heterogeneity by tuning the $p$ factor in synthetic networks to reduce the number of necessary measurements for observability and further increase the contraction size providing more possible countermeasures for observability recovery.

Note that the above mentioned results are applicable for specific networks. In other words, we claim the results regarding the size/distribution of contractions and the clustering coefficient only for power-law degree distribution networks (SF-CSF networks). Further, the results on the relation of degree-heterogeneity and size/distribution of contractions are only stated for networks with Small-World property. For other kind of networks, for example Erdos-Renyi graphs, such results may not apply in general. Note that to make a justified claim about effect of clustering-coefficient/degree-heterogeneity we need to keep other graph properties (e.g. degree distribution, average degree, number of edges) unchanged so we can claim that the only effective property is clustering-coefficient/degree-heterogeneity. We cannot claim this for general graphs as they may differ in terms of, for example, degree distribution.

It should be noted that the algorithms to check the matching properties of the network, namely Hopcroft-Karp algorithm \cite{hopcraft} or the Dulmage-Mendelsohn decomposition \cite{dulmage58} are of $\mc{O}(n^{2.5})$ complexity. Particularly, the complexity of Algorithm~\ref{alg_cont} is in polynomial order $\mc{O}(n^{2.5})$. Note that, polynomial time algorithms are suitable for large-scale system analysis as their running time is upper-bounded by a polynomial expression in system size.
The polynomial order complexity of the algorithms  motivates application in observability analysis of large-scale networks/systems similar to the real examples given in the previous section. It is worth mentioning that,   the results in this paper can be extended to the dual case of large-scale network controllability.

As the final comment, it should be noted that this paper considers undirected networks and system graphs. The reason is that for directed networks \textit{root SCCs} play important role in observability \cite{liu_pnas,jstsp14}. Therefore, along with contractions, root SCCs are effective in observability recovery. In order to solely consider the role of contractions in observability recovery in this paper we focus on undirected networks. As the direction of future research, we plan to seek whether other graph properties such as network community structure and degree-degree correlation \cite{posfai2013correlation} are effective on the contraction analysis and observability properties.

\section*{Acknowledgement}
The authors would like to thank Professor Glenn Lawyer from Max-Planck-Institute for Informatics for providing us with real network data.

\bibliographystyle{IEEEbib}
\bibliography{bibliography}
%
%

%

\begin{IEEEbiography}[{\includegraphics[width=1.1in]{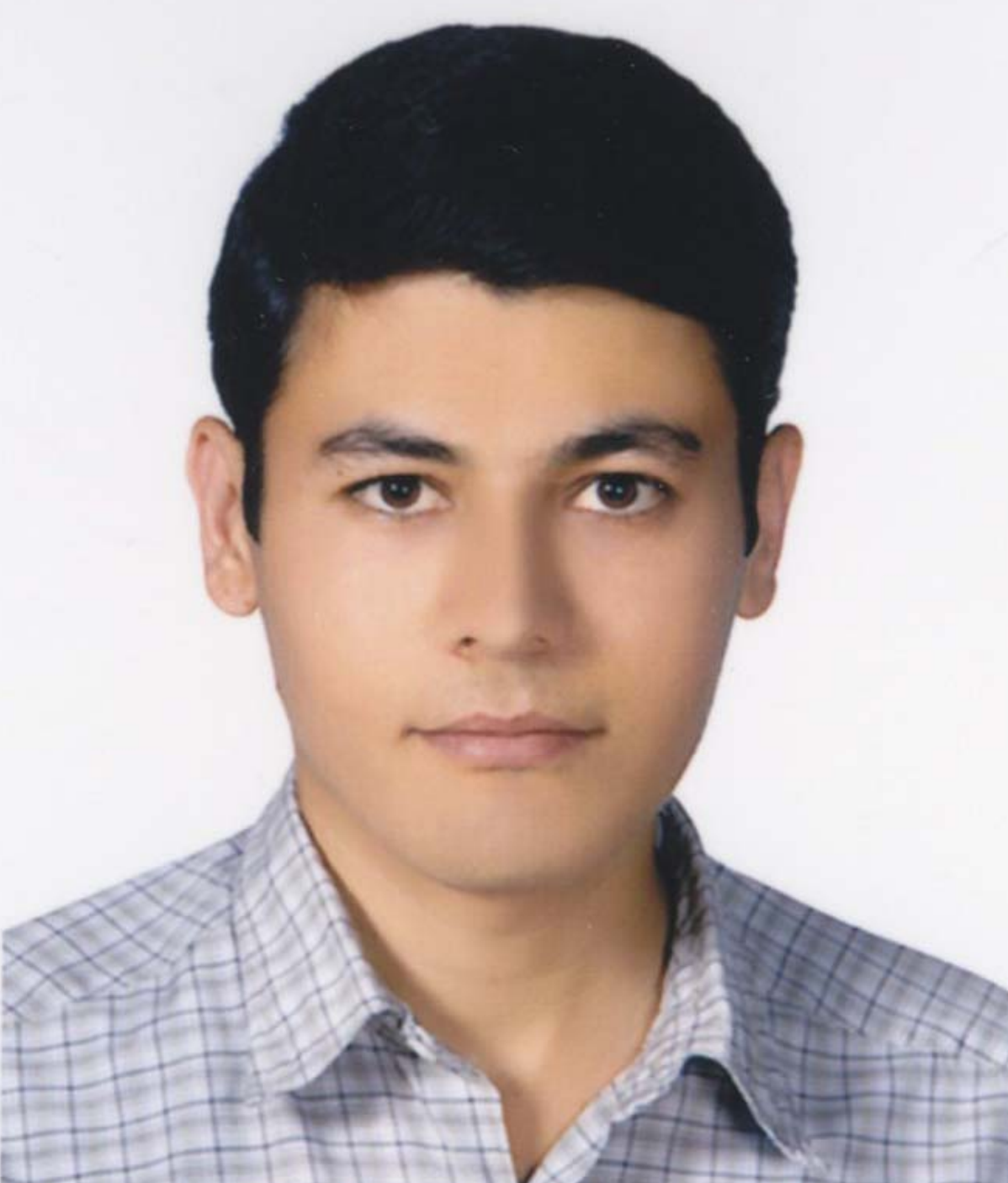}}]{Mohammadreza~Doostmohammadian}
received his B.Sc. and M.Sc. in Mechanical Engineering from Sharif University of Technology, Tehran, Iran, respectively in 2007 and 2010, where he worked on different applications of control systems and robotics. He received his PhD in Electrical and Computer Engineering from Tufts University, Medford, USA in 2015. During his PhD at Signal Processing and Robotic Network (SPARTN) lab he worked on control and signal processing over networks with particular application in social networks. From 2015 to 2017 he was a post-doctoral researcher at ICT Innovation Center for Advanced Information and Communication Technology (AICT), School of Computer Engineering, Sharif University of Technology, where he focused his research on control and estimation over networks and network epidemic. He is currently a researcher at Iran Telecommunication Research Center (ITRC), Tehran, Iran and a lecturer at Semnan University, Semnan, Iran. His general research interests include control and estimation, complex networks, and graph theory. He is a reviewer for IFAC and IEEE journals and conferences.
\end{IEEEbiography}

\begin{IEEEbiography}[{\includegraphics[width=1.1in]{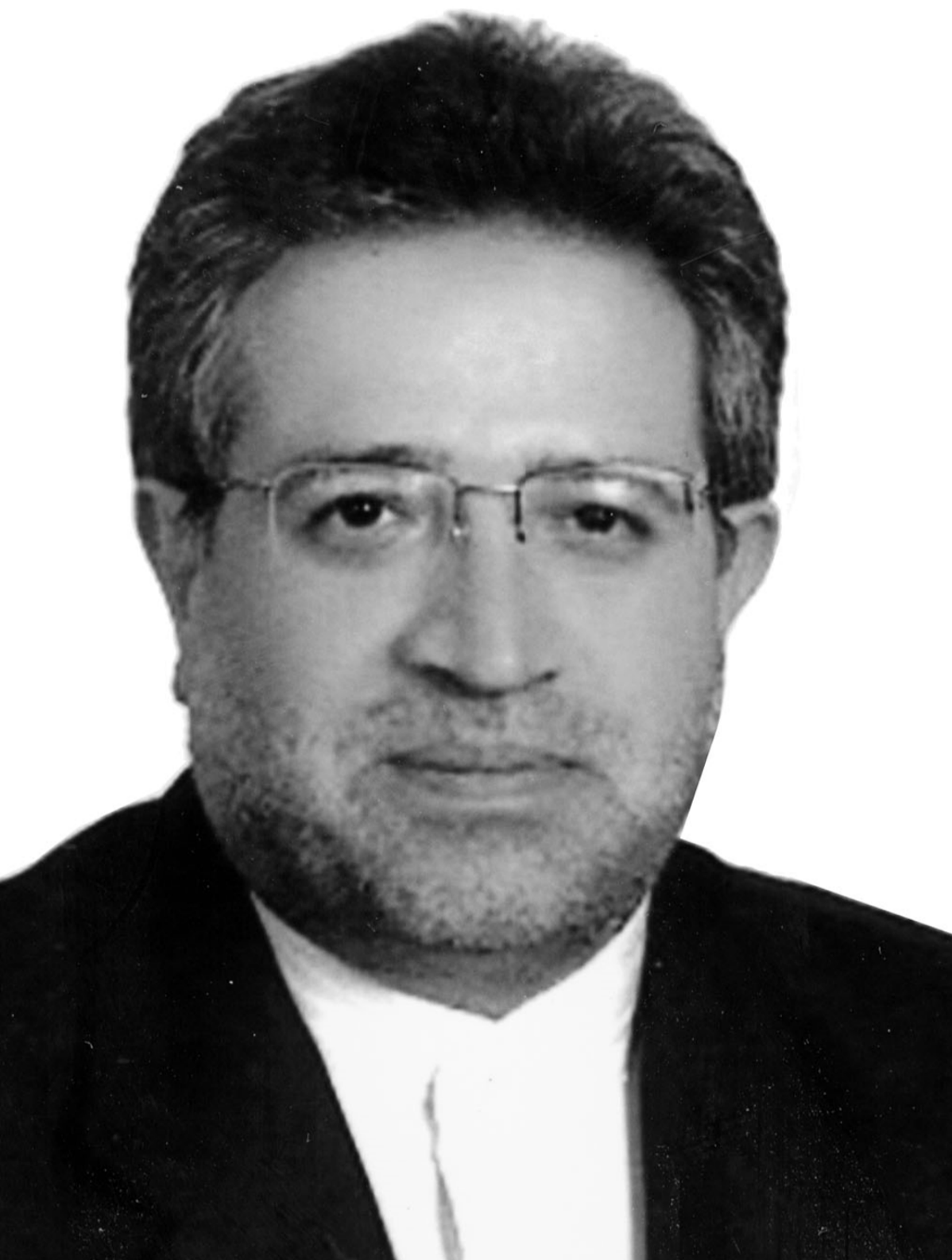}}]{Hamid R. Rabiee}
	received his B.S. and M.S. degrees (with great distinction) in electrical engineering from California State University, Long Beach, CA, in 1987 and 1989, respectively; the EEE degree in electrical and computer engineering from the University of Southern California, Los Angeles, CA, in 1993; and the Ph.D. degree in electrical and computer engineering from Purdue University, West Lafayette, IN, in 1996. From 1993 to 1996, he was a Member of Technical Staff at AT\&T Bell Laboratories. From 1996 to 1999, he worked as a Senior Software Engineer at Intel Corporation. From 1996 to 2000, he was also an Adjunct Professor of electrical and computer engineering at Portland State University, Portland, OR; Oregon Graduate Institute, Beaverton, OR; and Oregon State University, Corvallis, OR. Since September 2000, he has been with the Department of Computer Engineering, Sharif University of Technology, Tehran, Iran, where he is currently a Professor of computer engineering and Director of Sharif University Advanced Information and Communication Technology Research Institute (AICT), Digital Media Laboratory (DML), and Mobile Value Added Services Laboratory (M-VASL). He was also the founder of AICT, Advanced Technologies Incubator (SATI), DML, and M-VASL. He has been the Initiator and Director of national- and international-level projects in the context of United Nation Open Source Network program and Iran National ICT Development Plan, and holds 3 patents. Prof. Rabiee has received numerous awards and honors for his industrial, scientific, and academic contributions. He is a Senior Member of IEEE.
\end{IEEEbiography}

\begin{IEEEbiography}[{\includegraphics[width=1.1in]{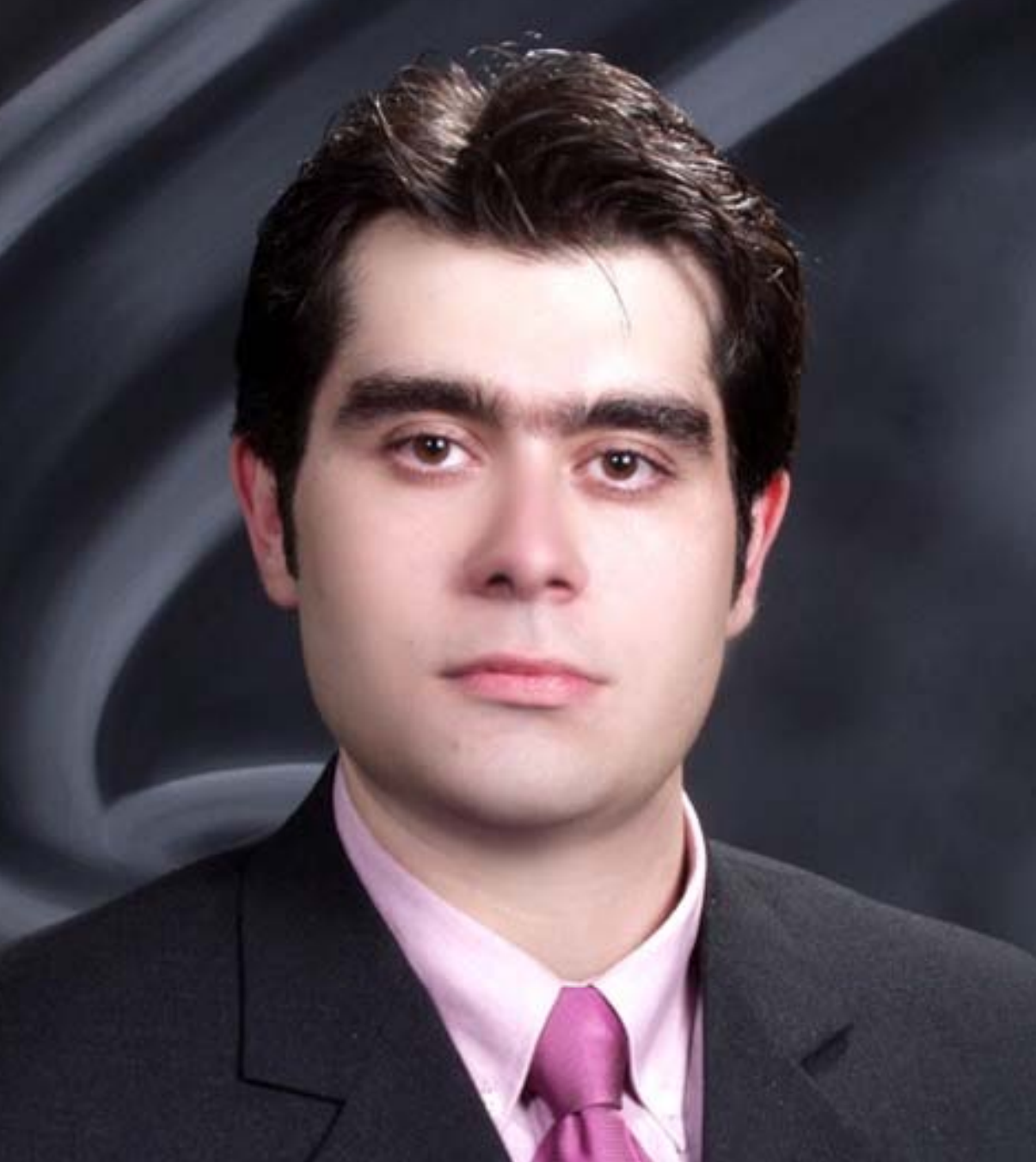}}]{Houman Zarrabi}
	received his PhD from Concordia University in Montreal, Canada in 2011. Since then he has been involved in various industrial and research projects. His main expertise includes IoT, M2M, CPS, big data, embedded systems and VLSI. He is currently the national IoT program director and assistant professor at Iran Telecommunication Research Center (ITRC). 
\end{IEEEbiography}

\begin{IEEEbiography}[{\includegraphics[width=1.1in]{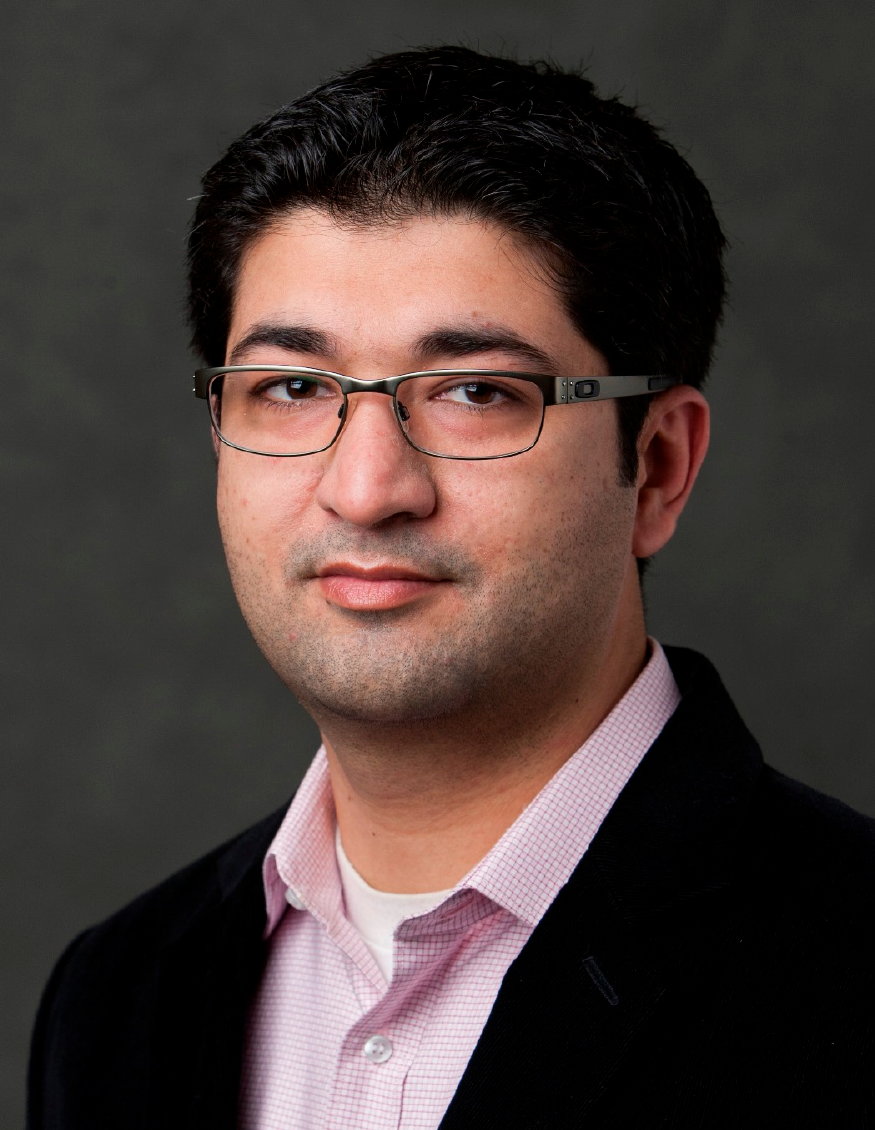}}]{Usman Khan}
	received his B.S. degree (with honors) in Electrical Engineering from University of Engineering and Technology,
	Lahore-Pakistan, in 2002, M.S. degree in Electrical
	and Computer Engineering (ECE) from the University
	of Wisconsin-Madison (UW-Madison) in 2004,
	and Ph.D. degree in ECE from Carnegie Mellon
	University in 2009. Currently, he is an Assistant Professor
	with the ECE Department at Tufts University.
	He received the NSF Career award in January 2014
	and was elevated to IEEE Senior Member grade
	in March 2014. His research interests include statistical signal processing, 	networked control and estimation, and distributed linear/nonlinear iterative
	algorithms for efficient operation and planning of complex infrastructures.
	
	Prof. Khan was a post-doctoral researcher in the Electrical and Systems Engineering department at the University of Pennsylvania from September 2009 to December 2010. He worked as a researcher in National Magnetic Resonance
	Facility at Madison (NMRFAM) from 2003 to 2005, as a research assistant in the Computer Science Dept. at UW-Madison from 2004 to 2005, and as an intern in AKAMAI Technologies in 2007. Dr. Khan is an associate
	member of the Sensor Array and Multichannel Technical Committee with the IEEE Signal Processing Society. He served on the Technical Program Committees
	of several IEEE conferences and has organized and chaired several IEEE workshops and sessions. His work was presented as Keynote speech at BiOS SPIE Photonics West-Nanoscale Imaging, Sensing, and Actuation for Biomedical
	Applications IX. 
\end{IEEEbiography}

%

\end{document}